\documentclass[twocolumn,tighten,times]{aastex63}

\usepackage{xspace}
\usepackage{booktabs}
\usepackage{lmodern}
\usepackage{slantsc}

\newcommand{\gaia}{{\em Gaia}\xspace}
\newcommand{\tess}{{\em TESS}\xspace}
\newcommand{\wise}{{\em WISE}\xspace}
\newcommand{\galex}{{\em GALEX}\xspace}
\newcommand{\obj}{ZTF\,J0328$-$1219\xspace}
\newcommand{\teff}{$T_{\mathrm{eff}}$\xspace}
\newcommand{\logg}{$\log(g)$\xspace}
\newcommand{\calcium}{[Ca/He]\xspace}
\newcommand{\hydro}{[H/He]\xspace}
\newcommand{\Aperiod}{9.9370\xspace}

\received{March 20, 2021}
\revised{May 25, 2021}
\accepted{June 2, 2021}
\submitjournal{ApJ}

\shorttitle{ZTF\,J0328-1219: Recurring Transits and Circumstellar Gas}
\shortauthors{Vanderbosch et al.}
\graphicspath{{./}{figures/}}

\begin{document}

\title{Recurring Planetary Debris Transits and Circumstellar Gas around White Dwarf ZTF\,J0328--1219}


\correspondingauthor{Zach Vanderbosch}
\email{zvanderbosch@astro.as.utexas.edu}

\author[0000-0002-0853-3464]{Zachary P. Vanderbosch}
\affiliation{Department of Astronomy, University of Texas at Austin, Austin, TX-78712, USA}
\affiliation{McDonald Observatory, Fort Davis, TX-79734, USA}

\author[0000-0003-3182-5569]{Saul Rappaport}
\affiliation{Department of Physics and Kavli Institute for Astrophysics and Space Research, Massachusetts Institute of Technology, Cambridge, MA 02139,
USA}

\author[0000-0001-9632-7347]{Joseph A. Guidry}
\affiliation{Department of Astronomy, University of Texas at Austin, Austin, TX-78712, USA}

\author[0000-0002-4080-1342]{Bruce L. Gary}
\affiliation{Hereford Arizona Observatory, Hereford, AZ 85615, USA}

\author[0000-0002-9632-1436]{Simon Blouin}
\affiliation{Los Alamos National Laboratory, PO Box 1663, Los Alamos, NM 87545, USA}

\author[0000-0001-7996-618X]{Thomas G. Kaye}
\affiliation{Foundation for Scientific Advancement, Sierra Vista, AZ, USA}

\author[0000-0001-6654-7859]{Alycia J. Weinberger}
\affiliation{Earth and Planets Laboratory, Carnegie Institution for Science, 5241 Broad Branch Rd NW, Washington, DC 20015, USA}

\author[0000-0001-9834-7579]{Carl Melis}
\affiliation{Center for Astrophysics and Space Sciences, University of California, San Diego, CA 92093-0424, USA}

\author[0000-0001-5854-675X]{Beth L. Klein}
\affiliation{Department of Physics and Astronomy, University of California, Los Angeles, CA 90095-1562, USA}

\author[0000-0001-6809-3045]{B. Zuckerman}
\affiliation{Department of Physics and Astronomy, University of California, Los Angeles, CA 90095-1562, USA}

\author[0000-0001-7246-5438]{Andrew Vanderburg}
\affiliation{Department of Astronomy, University of Wisconsin-Madison, Madison, WI 53706, USA}

\author[0000-0001-5941-2286]{J.~J. Hermes}
\affiliation{Department of Astronomy, Boston University, 725 Commonwealth Ave., Boston, MA 02215, USA}

\author[0000-0003-4145-3770]{Ryan J. Hegedus}
\affiliation{University of North Carolina at Chapel Hill, Department of Physics and Astronomy, Chapel Hill, NC 27599, USA}

\author[0000-0003-0684-7803]{Matthew. R. Burleigh}
\affiliation{School of Physics and Astronomy, University of Leicester, Leicester, LE1 7RH, UK}

\author[0000-0003-3904-6754]{Ramotholo Sefako}
\affiliation{South African Astronomical Observatory, P.O Box 9, Observatory 7935, Cape Town, South Africa}

\author[0000-0003-0496-8732]{Hannah L. Worters}
\affiliation{South African Astronomical Observatory, P.O Box 9, Observatory 7935, Cape Town, South Africa}

\author[0000-0003-3868-1123]{Tyler M. Heintz}
\affiliation{Department of Astronomy, Boston University, 725 Commonwealth Ave., Boston, MA 02215, USA}

\begin{abstract}

We present follow-up photometry and spectroscopy of \obj strengthening its status as a white dwarf exhibiting transiting planetary debris. Using \tess and Zwicky Transient Facility photometry, along with follow-up high speed photometry from various observatories, we find evidence for two significant periods of variability at 9.937 and 11.2\,hr. We interpret these as most likely the orbital periods of different debris clumps. Changes in the detailed dip structures within the light curves are observed on nightly, weekly, and monthly timescales, reminiscent of the dynamic behavior observed in the first white dwarf discovered to harbor a disintegrating asteroid,  WD\,1145+017. We fit previously published spectroscopy along with broadband photometry to obtain new atmospheric parameters for the white dwarf, with $M_{\star} = 0.731 \pm 0.023\,M_{\odot}$, $T_{\mathrm{eff}} = 7630 \pm 140\,$K, and \calcium$=-9.55\pm0.12$. With new high-resolution spectroscopy, we detect prominent and narrow Na D absorption features likely of circumstellar origin, with velocities $21.4\pm1.0$ km~s$^{-1}$ blue-shifted relative to atmospheric lines. We attribute the periodically modulated photometric signal to dusty effluents from small orbiting bodies such as asteroids or comets, but are unable to identify the most likely material that is being sublimated, or otherwise ejected, as the environmental temperatures range from roughly 400\,K to 600\,K. 

\end{abstract}



\section{Introduction} 
\label{sec:intro}

White dwarfs, the most common endpoint of stellar evolution, are likely to host planetary systems which have survived post-main sequence evolution. Indirect evidence for the presence of these planets includes atmospheric metal pollution in 30\% of white dwarfs cooler than 20,000\,K \citep{Zuckerman2003,Zuckerman2010,Koester2014}, infrared excess caused by a warm circumstellar debris disc in more than 50 white dwarfs \citep{Zuckerman1987,Mullally2007,Dennihy_2020_1,Xu2020}, and in a small number of these objects, metallic emission lines due to a hot gaseous component of the debris disk \citep[][]{Gaensicke2006,Dennihy_2020_2,Melis2020,GF2020}. These observational signatures are generally agreed to be the result of the tidal disruption of a planetary body after being gravitationally perturbed onto a highly eccentric orbit by one or more of the surviving outer planets \citep{Debes2002,Jura2003}.

Another way to observe this planetary debris is via the transit method, when dust-emitting debris clumps passing in front of the white dwarf momentarily attenuate some of the stellar flux along our line of sight. To date, only two objects have been confirmed to exhibit recurring transit events due to planetary debris, WD\,1145+017 with debris in 4.5--4.9\,hr orbits \citep{Vanderburg2015}, and ZTF\,J0139+5245 with debris in a ${\approx}100$ day orbit \citep{Vanderbosch2020}. The more than 2-order of magnitude difference in orbital periods may be indicative of different evolutionary stages during the standard tidal disruption process, though other processes such as rotational fission \citep{Veras_2020_1} may be at play in ZTF\,J0139+5245. Regardless of the mechanism that is disrupting planetary bodies, it is likely that a broad distribution of orbital periods could be observable among this class of objects, and confirming more such objects is needed to empirically fill out and understand this distribution.

Now in the age of large scale time-domain surveys, the likelihood of finding new transiting planetary debris candidates has increased dramatically. Publicly available light curves from both current and future surveys such as the Zwicky Transient Facility \citep[ZTF,][]{Masci2019,Bellm2019}, the Catalina Real-Time Transient Survey \citep[CRTS,][]{Drake2009}, the Transiting Exoplanet Survey Satellite \citep[{\em TESS},][]{Ricker2014}, and the Vera C. Rubin Observatory's Legacy Survey of Space and Time \citep[LSST,][]{Ivezic2019}, can be assessed for variability and cross-matched with catalogs of known and candidate white dwarfs such as \gaia \citep[e.g.][]{GF2019} and the Montreal White Dwarf Database \citep[MWDD,][]{Dufour2017} to identify good candidates for transiting debris systems.

In a recent study by \citet{Guidry2021}, five new candidates for white dwarfs exhibiting variability caused by transiting planetary debris were identified using variability metrics based on \gaia DR2 and ZTF DR3 photometry. \citet{Guidry2021} obtained follow up spectroscopy and time series photometry to identify the spectral types of each object and confirm the presence of variability. One of these five objects, ZTF\,J032833.52$-$121945.27 (hereafter \obj), was identified as a metal-polluted, DZ white dwarf\footnote{DZ follows the white dwarf classification system of \citet{Sion1983} indicating a degenerate star whose dominant optical spectral features are from elements heavier than He.} with short-timescale, $\approx$10\,\% amplitude variability in its 3-hr light curve from McDonald Observatory. Given the promising evidence for transiting debris, we targeted this object for detailed follow-up observations.

In this work, we present new and archival observations and identify periodic optical variability and atomic absorption features likely of circumstellar origin in \obj, strengthening its classification as a white dwarf with transiting planetary debris. In Section~\ref{sec:obs} we describe the new observations obtained along with archival photometry, in Section~\ref{sec:results} we discuss the periodic variability, new white dwarf atmospheric parameters derived from photometric and spectroscopic observations, and the detection of circumstellar gas, in Section~\ref{sec:discussion} we discuss the possible nature of \obj and potential causes of the periodic variability, and in Section~\ref{sec:conclusions} we summarize our results and draw some conclusions.


\begin{deluxetable*}{rccccc|crccccc}
    \tablenum{1}
    \tablecaption{Journal of Time-Series Photometric Observations of ZTF\,J0328$-$1219}
    \label{tab:phot_obs}
    \tabletypesize{\footnotesize}
    \tablewidth{0pt}
    \tablehead{
    \colhead{Date} & \colhead{Start} & \colhead{Telescope} & \colhead{Filter} & \colhead{$t_\mathrm{exp}$} & \colhead{Length} && 
    \colhead{Date} & \colhead{Start} & \colhead{Telescope} & \colhead{Filter} & \colhead{$t_\mathrm{exp}$} & \colhead{Length}\\
    \colhead{} & \colhead{(UTC)} & \colhead{} & \colhead{} & \colhead{(s)} & \colhead{(hr)} &&
    \colhead{} & \colhead{(UTC)} & \colhead{} & \colhead{} & \colhead{(s)} & \colhead{(hr)}
    }
    \startdata
     2020 Oct 19 & 06:30:09  & McD  & BG40  & 10 & 3.19   &&   2021 Jan 03 & 02:19:14  & JBO  & clear & 60 & 4.88 \\
     2020 Dec 02 & 06:16:58  & HAO  & clear & 90 & 4.29   &&   2021 Jan 05 & 01:58:16  & JBO  & clear & 60 & 5.82 \\
     2020 Dec 03 & 01:32:32	 & HAO  & clear & 90 & 9.01   &&   2021 Jan 06 & 02:09:04  & JBO  & clear & 60 & 6.23 \\
     2020 Dec 04 & 04:38:48	 & HAO  & clear & 90 & 5.88   &&   2021 Jan 08 & 01:38:02  & JBO  & clear & 60 & 5.90 \\
     2020 Dec 05 & 01:33:34	 & HAO  & clear & 90 & 8.92   &&   2021 Jan 12 & 01:39:36  & HAO  & clear & 90 & 5.80 \\
     2020 Dec 06 & 01:28:18	 & HAO  & clear & 90 & 8.89   &&   2021 Jan 13 & 01:38:10  & HAO  & clear & 90 & 5.69 \\
     2020 Dec 07 & 02:42:08	 & HAO  & clear & 90 & 7.35   &&   2021 Jan 14 & 01:38:43  & HAO  & clear & 90 & 5.69 \\
     2020 Dec 12 & 01:25:19	 & HAO  & clear & 90 & 3.49   &&   2021 Jan 15 & 01:48:57  & HAO  & clear & 90 & 5.60 \\
                 & 03:22:49  & McD  & BG40  & 10 & 5.59   &&   2021 Jan 16 & 01:38:55  & HAO  & clear & 90 & 5.75 \\
     2020 Dec 13 & 01:30:24	 & HAO  & clear & 90 & 8.40   &&   2021 Jan 17 & 01:39:15  & HAO  & clear & 90 & 4.91 \\
                 & 02:45:03	 & McD  & BG40  & 10 & 5.90   &&   2021 Jan 18 & 01:42:10  & HAO  & clear & 90 & 5.47 \\
                 & 19:08:02	 & SAAO & V     & 60 & 3.05   &&   2021 Jan 27 & 01:47:34  & HAO  & clear & 90 & 4.79 \\
     2020 Dec 14 & 02:36:19	 & McD  & BG40  & 20 & 6.56   &&   2021 Jan 31 & 01:43:18  & HAO  & clear & 90 & 4.56 \\
     2020 Dec 15 & 01:33:39	 & HAO  & clear & 90 & 8.29   &&   2021 Feb 01 & 01:43:19  & HAO  & clear & 90 & 2.93 \\
                 & 02:32:49	 & McD  & BG40  & 20 & 6.63   &&   2021 Feb 05 & 01:52:07  & HAO  & clear & 90 & 4.12 \\
     2020 Dec 16 & 01:58:31	 & HAO  & clear & 90 & 7.87	  &&   2021 Feb 06 & 01:54:34  & HAO  & clear & 90 & 3.88 \\
     2020 Dec 17 & 01:27:23	 & HAO  & clear & 90 & 8.43   &&   2021 Feb 07 & 01:52:29  & HAO  & clear & 90 & 4.04 \\
                 & 03:53:41	 & JBO  & clear & 60 & 5.40   &&   2021 Feb 09 & 01:31:56  & McD  & BG40  & 5  & 3.76 \\
     2020 Dec 20 & 02:59:04	 & JBO  & clear & 60 & 6.24   &&               & 01:51:44  & HAO  & clear & 90 & 4.03 \\
     2020 Dec 21 & 04:50:08	 & JBO  & clear & 60 & 3.99   &&   2021 Feb 10 & 01:26:58  & McD  & BG40  & 5  & 3.82 \\
     2020 Dec 22 & 04:24:24	 & JBO  & clear & 60 & 4.02   &&   2021 Feb 11 & 01:35:51  & McD  & BG40  & 10 & 3.63 \\
     2020 Dec 31 & 04:40:12	 & JBO  & clear & 60 & 3.59   &&   2021 Feb 12 & 01:28:04  & McD  & BG40  & 10 & 3.80 \\
     2021 Jan 02 & 05:16:32  & JBO  & clear & 60 & 3.02   &&               & 01:55:02  & HAO  & clear & 90 & 3.90 \\
    \enddata
\end{deluxetable*}

\section{Observations} 
\label{sec:obs}

\subsection{Time-Series Photometry}

\textit{McDonald/ProEM}: We acquired high-speed time-series photometry of \obj on nine nights, 2020 October 19, 2020 December 12--15, and 2021 February 9--12, with the McDonald Observatory (McD) 2.1-m Otto Struve telescope using the Princeton Instruments ProEM frame-transfer CCD. Located at Cassegrain focus, the camera has a $1.6\times1.6$ arcmin field of view. We used $4\times4$ binning resulting in 0.38\arcsec\ per binned pixel. We exclusively used the blue-bandpass BG40 filter, with exposure times of 5, 10 or 20-s. Using standard calibration images acquired each night, we bias, dark, and flat field corrected each science exposure with standard {\sc iraf} routines. We performed circular aperture photometry on \obj and the one available comparison star in our field of view using the {\sc iraf} package {\sc ccd\_hsp} \citep{Kanaan2002}, with aperture radii ranging from 1.5 to 10 pixels in half-pixel steps. 

We extracted light curves generated from each aperture radius using the {\sc phot2lc}\footnote{\href{https://phot2lc.readthedocs.io/en/latest/?badge=latest}{phot2lc.readthedocs.io}} python package, and performed differential photometry to remove atmospheric effects. With {\sc phot2lc} we clipped $\ge 5\sigma$ outliers within a sliding window of width 30-min, and divided each light curve by a low-order polynomial of degree two or less to remove airmass trends. We selected the optimal aperture size based on the light curve with the lowest average point-to-point scatter. Lastly, we used Astropy \citep{Astropy_2018,Astropy_2013} within {\sc phot2lc} to apply barycentric corrections to the mid-exposure time stamps of each image. A summary of McDonald observations can be found in Table~\ref{tab:phot_obs}.

\textit{HAO}: We acquired time-series photometry on 26 nights between 2020 December 2--17 (10 nights), 2021 January 12--18 (7 nights), and 2021 January 27 -- February 12 (8 nights) using the Astro-Tech 16-in Ritchey-Chretien telescope located at Hereford Arizona Observatory (HAO). We used an SBIG ST-10XME CCD camera with a KAF 3200E main chip to acquire a series of 90-s exposures with 9-s readout times each night using a clear filter. We performed bias, dark, and flat-field calibrations, along with circular aperture photometry, using the Maxim DL program. We tested aperture radii of 4, 5, and 6 pixels on the target and comparison stars, choosing a weighted-average of the light curves corresponding to apertures which maximized the signal-to-noise of each night's observations. Six comparison stars were used to correct for atmospheric effects and calibrate the brightness of ZTF\,J0328-1219 using magnitudes and B--V colors from the AAVSO Photometric All-Sky Survey (APASS) archive. We applied extinction corrections to our target as a function of airmass, using an empirically determined correction function based on our selection of comparison stars. Lastly, we applied barycentric timestamp corrections using Jason Eastman's Barycentric Julian Date web page\footnote{\url{http://astroutils.astronomy.ohio-state.edu/time/utc2bjd.html}}. A summary of HAO observations can be found in Table~\ref{tab:phot_obs}.

\begin{deluxetable*}{cllccclr}
\tablenum{2}
\tablecaption{Journal of Spectroscopic Observations of ZTF\,J0328$-$1219}
\label{tab:spec_obs}
\tabletypesize{\footnotesize}
\tablewidth{0pt}
\tablehead{
    \colhead{Date} & 
    \colhead{Facility/Instrument} &  
    \colhead{$t_\mathrm{exp}$} & 
    \colhead{Seeing} &
    \colhead{Airmass} &
    \colhead{Slit Width} &
    \colhead{Wavelengths} & 
    \colhead{$R$}\\
    \colhead{} & \colhead{} & \colhead{(s)} & \colhead{(\arcsec)} &
    \colhead{} & \colhead{(\arcsec)} & \colhead{(\AA)} & \colhead{($\lambda/\Delta\lambda$)}
}
\startdata
 2020 Nov 16 & LDT/DeVeny    & 6$\times$180   & 1.7 & 1.50 & 1.0 & 3700$-$6800               & 1200  \\
 2020 Dec 07 & Magellan/MIKE & 2$\times$1800  & 0.5 & 1.06 & 1.0 & 3500$-$5060B$^{\dagger}$  & $28{,}000$ \\
             &               &                &     &      &     & 5000$-$9400R$^{\dagger}$  & $22{,}000$ \\
 2020 Dec 21 & SOAR/Goodman  & 7$\times$600   & 1.2 & 1.16 & 1.0 & 7560$-$8750               & 3000  \\ 
\enddata
\tablenotetext{\dagger}{B and R correspond to the blue and red arms of the MIKE spectrograph.}
\end{deluxetable*}

\textit{JBO}: We acquired time-series photometry on 10 nights between 2020 December 16 and 2021 January 7, using a 32-in Ritchey-Chretien telescope located at Junk Bond Observatory (JBO). On each night we acquired a series of 60-s exposures without any filter. We performed bias, dark, and flat-field corrections, along with circular aperture photometry, using the AstroImageJ program \citep{Collins2017}. We performed differential photometry using the same six comparison stars used for the HAO observations, but no detrending procedures were applied. We performed barycentric corrections to the mid-exposure timestamps of each image using Astropy. A summary of JBO observations can be found in Table~\ref{tab:phot_obs}.

\textit{SAAO}: We acquired time-series photometry on one night, 2020 December 13, with the South African Astronomical Observatory (SAAO)'s 1-m telescope called Lesedi, and the Sutherland High-speed Optical Camera \citep[SHOC;][]{Coppejans_2013}. The target was observed continuously for three hours with 60-s exposures in V band. $2\times2$ binning was used, giving $0.67$~arcsecs per binned pixel. The field of view of the SHOC camera on Lesedi is $5.7\times5.7$~arcmin~squared, allowing several comparison stars of similar brightness to be monitored for differential photometry. Conditions were very good, with seeing $\approx1.2$\arcsec. The images were bias and flat-field corrected using the local SAAO SHOC pipeline, which is driven by python scripts running {\sc iraf} tasks (pyfits and pyraf). Aperture photometry was performed using the Starlink package {\sc autophotom}. We used a 3 pixel radius aperture that maximized the signal to noise, and the background was measured in an annulus surrounding this aperture. Differential photometry was then performed using three comparison stars. The resultant light curve was detrended for air mass effects with a second-order polynomial, and a barycentric correction was applied to the time stamps using Astropy routines. A summary of SAAO observations can be found in Table~\ref{tab:phot_obs}.

\textit{TESS}: \obj (TIC\,93031595) was observed by \tess in Sector 4 (2018 October 18 -- November 14) and in Sector 31 (2020 October 22 -- November 18). During Sector 4, \obj was not specifically targeted for observations, so the pixels surrounding \obj were only downloaded after being co-added to an exposure time of 30 minutes. By the time target lists were assembled for Sector 31, however, \obj had been identified as a white dwarf candidate by \textit{Gaia} DR2 \cite{GF2019}, so it was targeted with \tess's two-minute cadence mode. The Sector 4 and 31 light curves have durations of 25.8 and 25.4\,days, and data gaps lasting 6.2 and 2.2\,days, respectively. 

We extracted the Sector 4 \tess light curves with a custom aperture photometry pipeline. We downloaded the pixels surrounding \obj\ using the TESSCut interface at the Mikulski Archive for Space Telescopes and extracted a light curve from a single-pixel stationary aperture (the closest pixel to the R.A. and Decl. of \obj). We corrected for spacecraft systematics and scattered light by decorrelating the light curve against the mean and standard deviation of the spacecraft quaternion time series within each exposure \citep[see][]{vanderburg2019} and the time series of background flux outside the photometric aperture. Since \obj was observed with two-minute cadence observations in Sector 31, its light curve was extracted by the official Science Processing Operation Center (SPOC) pipeline \citep{Jenkins:2015, Jenkins:2016} based at NASA Ames Research Center. The SPOC pipeline works by performing simple (stationary) aperture photometry and removing common-mode systematic errors with its Pre-search Data Conditioning module \citep{smith, stumpe}.

\textit{ZTF}: We queried the public ZTF survey \citep{Masci2019,Bellm2019} for $g$ and $r$-band DR5 photometry using the NASA/IPAC Infared Science Archive (IRSA). Following recommendations in the ZTF Science Data System Explanatory Supplement\footnote{\url{http://web.ipac.caltech.edu/staff/fmasci/ztf/ztf_pipelines_deliverables.pdf}}, we removed poor quality detections by requiring $\textit{catflags} = 0$ and $|\textit{sharp}| < 0.5$. The resulting ZTF light curve has 348 data points ($g=166$, $r=182$), all with exposure times of 30-s and spanning 898 days between 2018 August 4 and 2021 January 18. Following light curve filtering, we applied barycentric corrections to the time stamps using Astropy.

\subsection{Spectroscopy}

\begin{deluxetable*}{lccDDDcrc}
\tablenum{3}
\tablecaption{MIKE Comparison Star Observations}
\label{tab:comp_obs}
\tabletypesize{\footnotesize}
\tablewidth{0pt}
\tablehead{
    \colhead{Target} & 
    \colhead{Source ID} &
    \colhead{Type} &
    \twocolhead{$G$} &
    \twocolhead{Distance} &
    \twocolhead{Sky Sep.$^{\dagger}$} &
    \colhead{Date Obs.} &
    \colhead{$t_\mathrm{exp}$} & 
    \colhead{Airmass} \\
    \colhead{} & \colhead{(Gaia eDR3)} & \colhead{} & \twocolhead{(mag)} & 
    \twocolhead{(pc)} & \twocolhead{(\arcmin)} &
    \colhead{} & \colhead{(s)} & \colhead{}
}
\decimals
\startdata
HD\,22243          & 5163917352681876736 & A0  & 6.23  & 118.3 & 172.5  & 2020 Dec 08  & 36  & 1.06 \\
HD\,21917          & 5161837970035375488 & G6V & 8.75  & 73.5  &  45.0  & 2020 Dec 08  & 96  & 1.05 \\
HD\,21634          & 5161919059017811072 & F3V & 8.71  & 132.6 &  17.3  & 2020 Dec 08  & 120 & 1.05 \\
Gaia\,J0328$-$1216 & 5161809451452454784 & --- & 12.15 & 122.2 &   3.7  & 2020 Dec 08  & 210 & 1.05 \\
\enddata
\tablecomments{Distances are the photogeometric values from \citet{Bailer-Jones_2021}, and Sky Sep. is the on-sky separation between each comparison star and \obj.}
\end{deluxetable*}

We detail here all of the spectroscopic observations obtained and used in this work. See Tables~\ref{tab:spec_obs} and \ref{tab:comp_obs} for a summary of the instrument and observational details.

\textit{SOAR/Goodman}: We carried out spectroscopic observations of \obj on 2020 December 21 using the Goodman Spectrograph \citep{Clemens2004} mounted on the 4.1-m Southern Astrophysical Research (SOAR) telescope. We used the Red Camera, the 1200l-R grating, a 1.0\arcsec\ long slit, and the GG-495 order-blocking filter to obtain a resolving power of $R{\approx}3000$ over the $7560-8750\,\mathrm{\AA}$ wavelength region. We obtained seven consecutive 600-s exposures, for a total exposure time of 1.17\,hr. We bias-subtracted, flat-fielded, wavelength-calibrated, and optimally extracted \citep{Marsh1989} each individual spectrum using a set of custom Python tools based on the {\sc zzceti\_pipeline}\footnote{\url{https://github.com/joshfuchs/ZZCeti_pipeline}} reduction package described in \citet{Fuchs2017}, before combining them into a final spectrum. We performed wavelength calibrations using the abundant night sky emission lines in this wavelength region, as identified by \citet{Osterbrock1996}. Lastly, we applied a heliocentric velocity correction to the combined spectrum.

\textit{LDT/DeVeny}: To obtain improved atmospheric parameters for \obj, we used the same spectrum presented in \citet{Guidry2021}, which was obtained using the DeVeny Spectrograph \citep[][]{Bida2014} mounted on the 4.3-m Lowell Discovery Telescope (LDT). On 2020 November 16 with $1.7\arcsec$ seeing, we acquired $6\times180$\,s exposures using the 300~line mm$^{-1}$ DV2 grating and a 1.0\arcsec\ slit to achieve a resolving power of $R{\approx}1200$ over 3700$-$6800\,\AA. We debiased and flat-fielded the exposures using standard {\sc starlink} routines \citep{Starlink}, and optimally extracted the spectrum \citep{Horne1986-1} using the software {\sc pamela}. Using {\sc molly} \citep{Marsh1989} we applied heliocentric velocity corrections to the wavelengths.

\textit{Magellan/MIKE}: We observed \obj on 2020 December 7 using the Magellan Inamori Kyocera Echelle (MIKE) spectrograph \citep{Bernstein2003} on the 6.5-m Magellan Clay Telescope at Las Campanas Observatory. Observations were done at an airmass of 1.06. Two consecutive integrations of 1800-s each were acquired with the 1$''$ wide slit for a resolving power of about $R{\approx}28{,}000$ in the blue channel (3500$-$5060\AA) and $R{\approx}22{,}000$ in the red channel (5000$-$9400\AA). The signal-to-noise (S/N) achieved was about 12 at 3960\AA, 27 at 5890\AA, and 30 at 6563\AA. Wavelength calibration was done with ThAr lamps taken just before and after the science exposures. The data were extracted, flat-fielded, and wavelength calibrated using the Carnegie Python MIKE pipeline which incorporates methods described in \citet{Kelson:2000} and \citet{Kelson:2003}. We applied Heliocentric velocity corrections calculated by the MIKE pipeline to each spectrum. On 2020 December 8, the same instrument setup was used to observe four stars near the line of sight to \obj to probe the interstellar medium (ISM). Details of the comparison stars and their observations can be found in Table~\ref{tab:comp_obs}.

\begin{deluxetable}{lllc}
\tablenum{4}
\tablecaption{Astrometric and Photometric Properties of \obj} 
\label{tab:phot_params}
\tabletypesize{\footnotesize}
\tablewidth{0pt}
\tablehead{\colhead{Parameter} & \colhead{Value} & \colhead{Uncertainty} & \colhead{Ref.}} 
\startdata
\gaia eDR3           & 5161807767825277184 & -- & 1 \\
R.A. (J2016)         & $03^{\mathrm{h}}28^{\mathrm{m}}33^{\mathrm{s}}.63$ & -- & 1 \\
decl. (J2016)        & ${-}12^{\circ}19'45''.49$   & -- & 1 \\
$\mu_{\mathrm{RA}}$ (mas yr$^{-1}$)  & 110.739     & 0.055 & 1 \\
$\mu_{\mathrm{dec}}$ (mas yr$^{-1}$) & ${-}$14.39  & 0.046 & 1 \\
parallax (mas)       & 23.038   & 0.056  & 1 \\
$d$ (pc)             & 43.32    & 0.12   & 2 \\
$G$                  & 16.6186  & 0.0035 & 1 \\
$G_{\rm{BP}}$        & 16.732   & 0.015  & 1 \\
$G_{\rm{RP}}$        & 16.446   & 0.011  & 1 \\
\textsl{\textsc{galex}} {\sc nuv} & 18.91    & 0.05   & 3 \\
$g$                  & 16.722   & 0.021  & 4 \\
$r$                  & 16.6750  & 0.0057 & 4 \\
$i$                  & 16.7754  & 0.0062 & 4 \\
$z$                  & 16.958   & 0.016  & 4 \\
$y$                  & 16.951   & 0.011  & 4 \\
$J$                  & 16.2969  & 0.0074 & 5 \\
$K_s$                & 16.188   & 0.035  & 5 \\
$W1$                 & 16.05    & 0.05   & 6 \\
$W2$                 & 15.88    & 0.10   & 6 \\
\enddata
\tablecomments{[1] \citet{Brown2020}, [2] \citet{Bailer-Jones_2021}, [3] \citet{Bianchi2017}, [4] \citet{Chambers2016}, [5] \citet{McMahon2013_VHS}, [6] \citet{Schlafly2019}}
\end{deluxetable}

\begin{figure*}[t!]
	\epsscale{1.15}
	\plotone{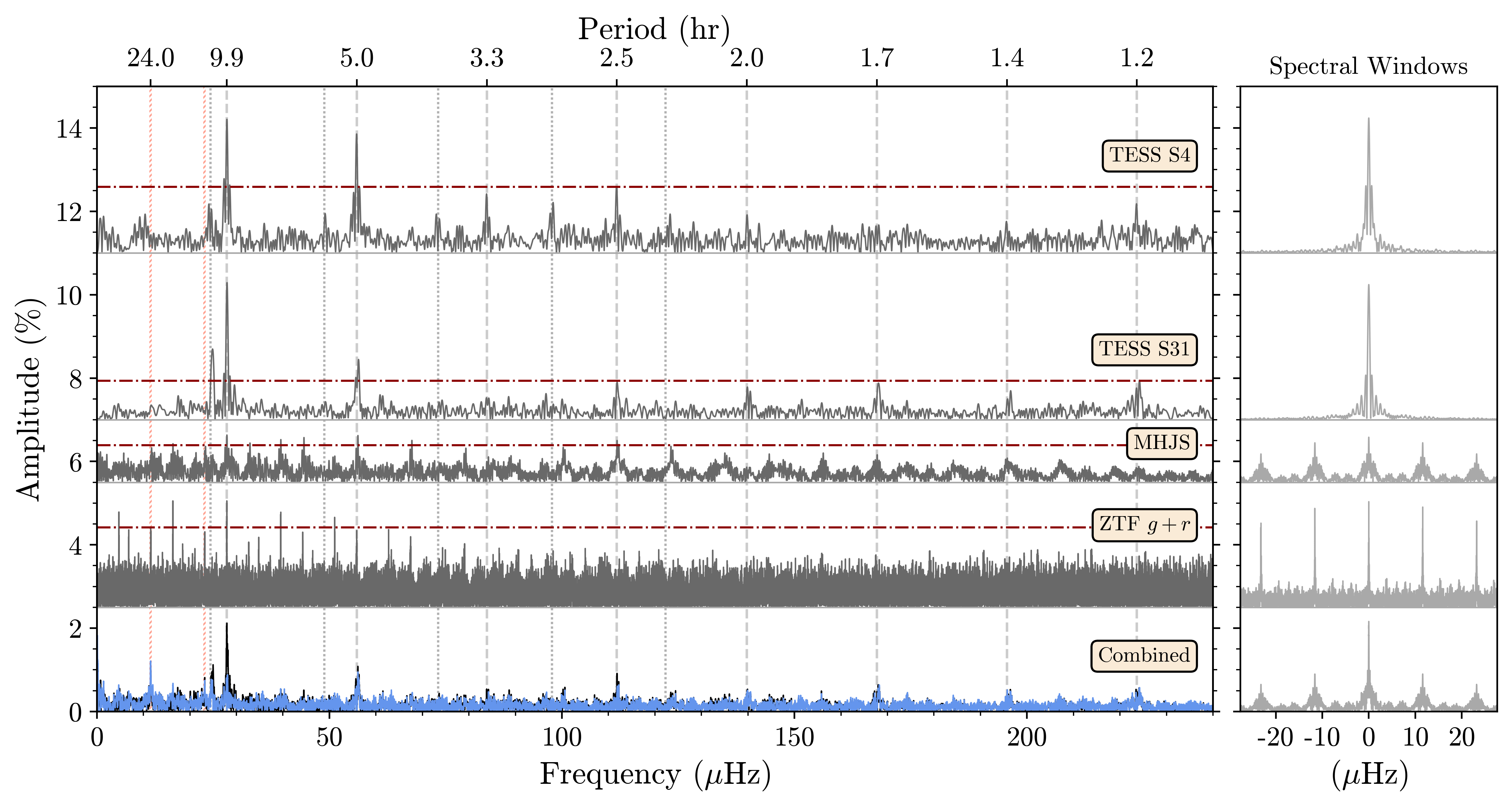}
	\caption{\textbf{Left}: Periodograms for different selections of time-series photometry, each vertically shifted for clarity, with the combined periodogram shown in black at bottom. MHJS stands for the combination of McDonald, HAO, JBO, and SAAO photometry, and horizontal red dash-dotted lines denote the 0.1\% false-alarm probability levels for each selection of data. The $P_A$ (9.937\,hr) period and its first seven harmonics are denoted with vertical dashed lines, while the vertical dotted lines denote the $P_B$ (11.2\,hr) period and its first four harmonics. Vertical red-hatched lines are placed at 11.6 and 23.2 $\mu$Hz to mark the 1-day and 0.5-day periods, respectively. Peaks at these frequencies only show up in the ground-based data, and we consider them to be artifacts caused by the diurnal sampling of those light curves. The over-plotted blue periodogram at bottom represents the combined data with the best-fit $P_{\mathrm{A}}$ and $P_{\mathrm{B}}$ values determined with {\sc Period04} prewhitened, along with 30 harmonics related to $P_{\mathrm{A}}$. Some significant residual power is still present after prewhitening, especially at harmonics of $P_{\mathrm{A}}$. For the MHJS periodogram, we excluded the single McDonald light curve from 2020 October 19 to reduce the amount of aliasing in the spectral window. \textbf{Right}: The spectral window of each data set, with the same x-axis scaling as the left panel, and amplitudes scaled to the highest peak in the corresponding periodogram. \label{fig:LSP}}
\end{figure*}

\vspace{1.0cm}
\subsection{SED Photometry}

To construct a spectral energy distribution (SED) for \obj, we retrieved photometry from the {\em Galaxy Evolution Explorer} (\galex, \citealt{Martin2005,Morrissey2005}), \gaia EDR3 \citep{Gaia1,Brown2020}, the Panoramic Survey Telescope and Rapid Response System \citep[Pan-STARRS1 or PS1,][]{Chambers2016,Magnier2013}, the VISTA Hemisphere Survey \citep[VHS,][]{McMahon2013_VHS}, and NASA's Wide-field Infrared Survey Explorer \citep[\wise,][]{Wright2010}. A summary of all the catalogued astrometry and photometry can be found in Table~\ref{tab:phot_params}. For the \wise W1 and W2 photometry, we used the values reported in the unWISE catalogue \citep{Schlafly2019} which uses deeper imaging and improved modeling of crowded regions compared to the ALLWISE catalogue \citep{Cutri2014}. For the distance, we used the {\sc photogeometric} value reported in \citet{Bailer-Jones_2021}.


\section{Results} \label{sec:results}

\begin{deluxetable*}{cc|ccccc|cccc}
\tablenum{5}
\tablecaption{Period Identifications} 
\label{tab:periods}
\renewcommand{\arraystretch}{1.1}
\tablecolumns{11}
\tablewidth{0pt}
\tabletypesize{\footnotesize}
\tablehead{\multicolumn{2}{c}{} & \multicolumn{5}{c}{{\sc Period04}} & \multicolumn{4}{c}{Gaussian Centroid} \\
          \cmidrule(lr){3-7}
          \cmidrule(lr){8-11}
          \colhead{Parameter} & \colhead{} & \colhead{ZTF$_{g{+}r}$} & \colhead{TESS-04} & 
           \colhead{TESS-31} & \colhead{MHJS$^{\dagger}$} & 
           \colhead{Combined} & \colhead{LS$^{\ddagger}$} & 
           \colhead{BLS$^{\ddagger}$} & \colhead{SH$^{\ddagger}$} &
           \colhead{Plavchan}
}
\startdata
$P_{\mathrm{A}}$                      & (hr) & $9.9374$    & $9.9314$    & $9.9243$    & $9.9324$    & $9.9370^*$   & $9.9246$    & $9.9225$    & $9.9161$    & $9.9191$    \\
$\sigma_{\mathrm{A}}^{**}$            & (hr) & $\pm0.0003$ & $\pm0.0074$ & $\pm0.0041$ & $\pm0.0012$ & $\pm0.0001$  & $\pm0.0002$ & $\pm0.0002$ & $\pm0.0003$ & $\pm0.0001$ \\
$P_{\mathrm{B}}$                      & (hr) & ---         & $11.549^*$  & $11.162^*$  & ---         & $11.1113$    & $11.1703$   & $11.1823$   & $11.1881$   & $11.1494$   \\
$\sigma_{\mathrm{B}}^{**}$            & (hr) & ---         & $\pm0.029$  & $\pm0.011$  & ---         & $\pm0.0002$  & $\pm0.0002$ & $\pm0.0006$ & $\pm0.0017$ & $\pm0.0001$ \\
HWHM$_{\mathrm{A}}^{\dagger\dagger}$  & (hr) & ---         & ---         & ---         & ---         & ---          & $0.052$     & $0.065$     & $0.057$     & $0.053$     \\
HWHM$_{\mathrm{B}}^{\dagger\dagger}$  & (hr) & ---         & ---         & ---         & ---         & ---          & $0.097$     & $0.111$     & $0.110$     & $0.093$     \\
\enddata
\tablenotetext{\dagger}{The combined McDonald, HAO, JBO, and SAAO data.}
\tablenotetext{\ddagger}{LS = Lomb-Scargle Periodogram, BLS = Box Least-Squares Periodogram, SH = Summed Harmonics FFT.}
\tablenotetext{$*$}{Values used to generate phase-folded light curves seen in Figures~\ref{fig:mcdphot}, \ref{fig:allphot}, and \ref{fig:Bfold}}
\tablenotetext{$**$}{\ Formal least squares uncertainties for the Period04 periods and for the centroids of the periodogram peaks.}
\tablenotetext{\dagger\dagger}{\ The measured half width at half maximum of the periodogram peaks which we take as a measure of the variations in the period and/or the photometric modulation profile with time.}
\end{deluxetable*}

\subsection{Periodic Variability} \label{subsec:period}

After the initial identification of ZTF\,J0328-1219 as a transiting planetary debris candidate due to its metal pollution and short-timescale variability \citep{Guidry2021}, we set out to investigate whether this object, like the two known dust cloud transiting systems WD\,1145+017 and ZTF\,J0139+5245, exhibits periodic variability related to the orbital period of planetary debris. The discovery of variability in \obj originally came from just 3\,hr of time-series photometry during a single night, during which no significant periodicity was observed. 

To identify possible periods on longer timescales, we first analyzed the light curves from \tess Sectors 4 and 31. We used Astropy routines to calculate unnormalized\footnote{See ``PSD normalization'' at the \href{https://docs.astropy.org/en/stable/timeseries/lombscargle.html}{Astropy Website}} Lomb-Scargle periodograms \citep[LS,][]{Lomb1976,Scargle1982} of the \tess light curves (see Figure~\ref{fig:LSP}), which we then converted from power ($p$) to fractional amplitude units ($A$) with the relation $A=\sqrt{4p/N}$, where $N$ is the total number of light curve data points. We estimated the false-alarm probability level of each periodogram using a bootstrap method \citep{Vanderplas2018,Bell2019}, where we randomly resampled each light curve 10,000 times with replacement, and with identical time samplings as the original light curves. In 99.9\% of cases, the maximum peak of the resulting periodograms did not exceed 1.59\% amplitude for Sector 4, and 0.85\% for Sector 31.

In both \tess sectors, a significant peak is present at around 27.9\,$\mu$Hz, or 9.93\,hr (hereafter labeled the A-period, $P_A$), along with several higher frequency peaks corresponding to harmonics of this period. In the \tess Sector 31 data, an additional significant peak at around 24.8\,$\mu$Hz, or 11.2\,hr (hereafter labeled the B-period, $P_B$) is present, and also appears in the Sector 4 data at lower significance. Some power appears to be present at the first four harmonics of the B-period in \tess Sector 4, but do not show up in \tess Sector 31.

Using the {\sc Period04} application \citep{Lenz2005}, we took the frequencies and amplitudes of the highest peaks near 9.93 and 11.2\,hr, and optimized them with a non-linear least squares fit of two independent sine curves to the light curves. We present the resulting periods and their formal least squares uncertainties \citep[$\sigma_A$ and $\sigma_B$,][]{Montgomery1999} in Table~\ref{tab:periods}. The resulting A-periods for the two sectors agree within 1-$\sigma$, while the B-periods exhibit a small but significant difference of 0.39\,hr, or about 9.7$\sigma$. We also examined the stability of the periods and amplitudes within each Sector by breaking the light curves up into five roughly 5-day segments and performing a least squares fit to each segment holding the A and B periods fixed at their best fit values for the whole sector. Both periods exhibit phase stability consistent with no change in period within each sector, though the B-period amplitude dropped significantly in Sector 31 between the first and last segment, from 2.1$\pm$0.3\% to 0.3$\pm$0.3\% respectively.

In Figure~\ref{fig:LSP}, we also show the periodograms for the combined McD, HAO, JBO, and SAAO data (labeled MHJS), for the combined ZTF $g+r$ data, and for the combination of all data sets presented in this work. For the MHJS periodogram, we only use data acquired from 2020 December through 2021 February to help reduce cycle count errors, which meant excluding the single McDonald run acquired in 2020 October. We use the same bootstrap method described above to define a 0.1\% false-alarm probability level for the ZTF data resulting in a 2.25\% threshold. For the multi-site MHJS data, however, we adopt four times the average periodogram amplitude between 0 and 300\,$\mu$Hz as an estimate of the 0.1\,\% false-alarm probability level \citep{Breger_1993_1,Kuschnig_1997_1}, resulting in a 0.93\,\% threshold. 

Both the ZTF and MHJS periodograms exhibit peaks aligned with the A-period and some of its harmonics. Several significant peaks at daily aliases of the A-period are also observed, which result from the spectral windows of the ZTF and MHJS light curves. Normally these aliases would make it difficult to determine the true period of variation, but the unambiguous detection of variability at 9.93\,hr by \tess helps identify the correct peaks in these ground-based observations. 

\begin{figure*}[t!]
	\epsscale{1.1}
	\plotone{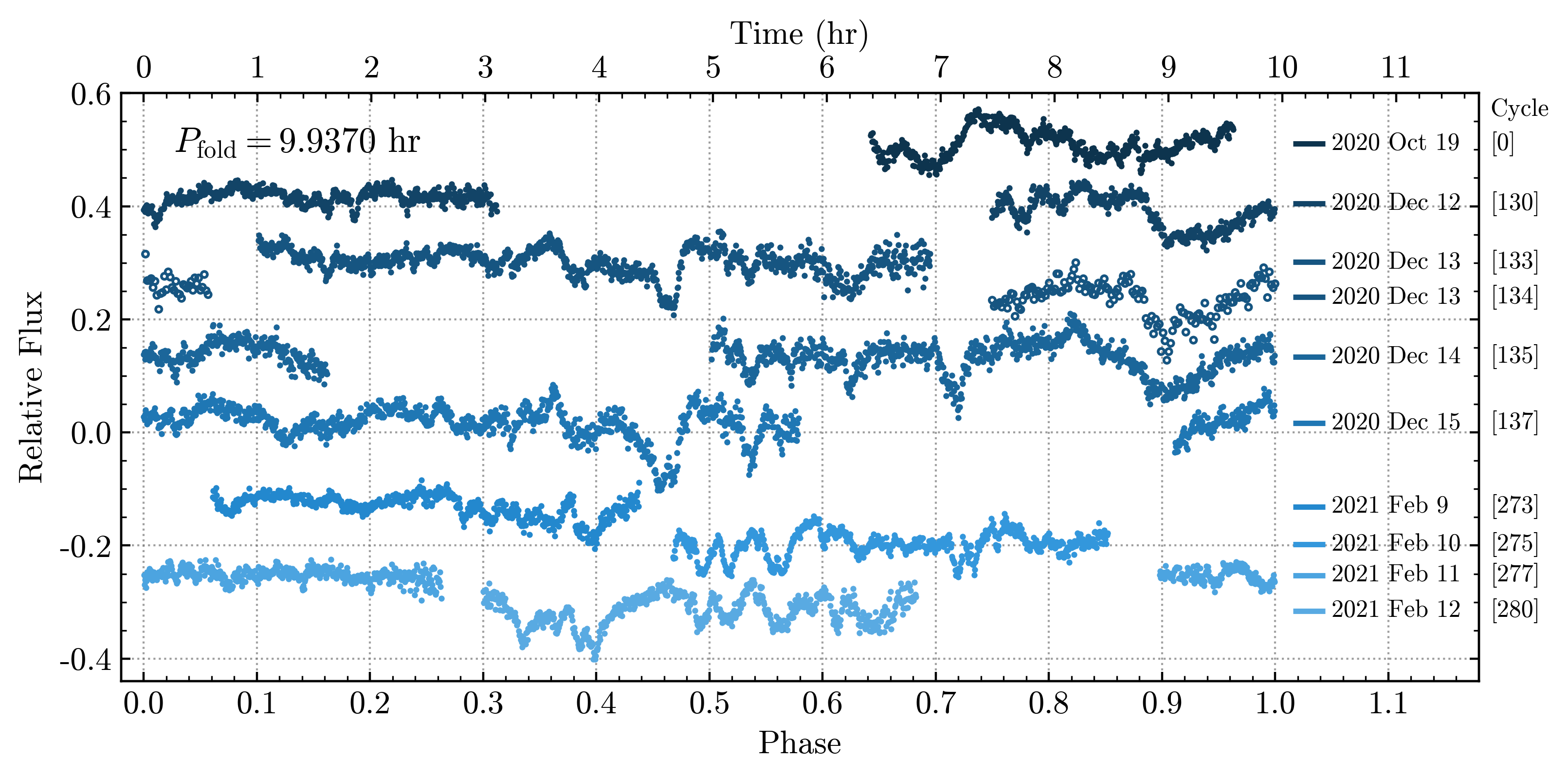}
	\caption{Individual light curves from McDonald Observatory (solid circles) and SAAO (open circles) phased to the best-fit $P_{\mathrm{A}}$ value of \Aperiod\,hr. Each successive light curve is vertically shifted down for easier comparison between nights and labeled by their corresponding dates. The number of $P_A$ cycles that have occurred since $\mathrm{BJD_{TDB}}=2459141.504938$ at the start of each night is given in brackets next to each date. \label{fig:mcdphot}}
\end{figure*}

We also searched for significant peaks corresponding to the B-period in the ZTF and MHJS periodograms. After prewhitening $P_A$ and 30 of its harmonics from the MHJS light curves, a small peak does show up near 23.5\,$\mu$Hz (11.8\,hr) just below the significance threshold. Given that $P_B$ changes by about 0.39\,hr between \tess Sectors 4 and 31, it is possible that this peak is also related to variability in \obj. However, this peak is also consistent with being an alias of a ${\approx}$1-day period, that may result from the daytime gaps in our ground-based observations coupled with slight differences in the mean flux level between nights and facilities. Peaks corresponding to a 1-day period (11.6\,$\mu$Hz) and its nearby daily alias (23.2\,$\mu$Hz) also show up quite prominently in the ground-based ZTF data (see peaks marked by vertical red-hatched lines in Figure~\ref{fig:LSP}). These peaks do not show up, however, in either of the \tess light curves, which do not suffer from daytime observational gaps. Thus, we consider any peaks that show up at the 1-day period, or aliases thereof, systematic artifacts caused by the diurnal sampling of our ground based data, and not related to variability within \obj.

We again used {\sc Period04} to optimize the periods for the ZTF, MHJS, and combined light curves. For ZTF and MHJS, we fit only a single sine curve for $P_A$ since the $P_B$ period was not significant in their periodograms. The resulting periods and formal least squares uncertainties are shown in Table~\ref{tab:periods}. These uncertainties are likely underestimated, as only the ZTF and \tess Sector 4 $P_A$ values agree within 3-sigma with the $P_A$ values determined from the combined data set. 

We also attempt period determinations with a separate method, using a Gaussian function to fit the peaks in the power spectra. This method was motivated by noticing that the observed periodogram peaks for the combined data set are significantly broader than would be expected for stable periodic variability, and that significant residual power remains near the A-period and its harmonics even after prewhitening the best-fit period and many of its harmonics from the data (see Figure~\ref{fig:LSP}). This suggests period or amplitude modulations are present in this system, or that multiple closely spaced periods exist near $P_A$ and $P_B$, any of which would make precise period determinations difficult for this system.

We only perform Gaussian fits for the combined data set, and use a non-linear least-squares routine to optimize the fit. We use the Gaussian center as a measure of the average period, and the Gaussian half-width-half-max (HWHM) as a measure of the effects that period modulations, amplitude modulations, or multiple closely space periods may have on the peak widths. We perform this fit on four different types of power spectra, a Lomb-Scargle periodogram, a box least-squares periodogram, a summed harmonics fast FT\footnote{For the summed-harmonics FFT, we assume that each frequency bin in a standard FT is the 10th harmonic of a corresponding base frequency, and we add the amplitudes of all 9 lower harmonics to what is already in that bin. This summation of the first 10 harmonics can be generalized to any choice of N harmonics.}, and a Plavchan periodogram \citep{Plavchan2008}, which may each be sensitive to different features in the light curves. The results of the fits are summarized in Table~\ref{tab:periods}, where we report the center and HWHM values for the Gaussian fits.

We also fit a Gaussian to the spectral window of the combined data set, and find that on average the spectral window is about three times narrower than the HWHM values for each period and for each type of power spectrum. The A-periods from the Gaussian fits appear shifted to lower values relative to the {\sc Period04} results, while the B-periods are closer to the {\sc Period04} results for \tess Sector 31 and the combined data set.

With the above analysis, we attempt to convey that given the evolution of the light curve modulation profile (see Section~\ref{subsec:lcs}), the possible changes in the periods over time, and the potential presence of multiple, closely-spaced periods, a precise period determination is difficult to obtain for \obj. Even if both $P_{\mathrm{A}}$ and $P_{\mathrm{B}}$ are highly stable over time, changes in the light curve modulation profile will themselves make precise determinations difficult. For simplicity, however, in the remainder of this work we use the best-fit $P_{\mathrm{A}}$ value determined using {\sc Period04} for the combined light curve, 9.9370\,hr, to construct phase-folded light curves and determine orbital parameters. Since $P_{\mathrm{B}}$ shows a significant change between the two \tess sectors, we use the best-fit periods determined with Period04 from each individual sector to generate their phase-folded light curves (11.549 and 11.162\,hr for Sectors 4 and 31, respectively, and opt to use the Sector 31 B-period to determine orbital parameters.

\subsection{Phase-Folded Light Curves} \label{subsec:lcs}

\begin{figure}[t!]
	\epsscale{1.1}
	\plotone{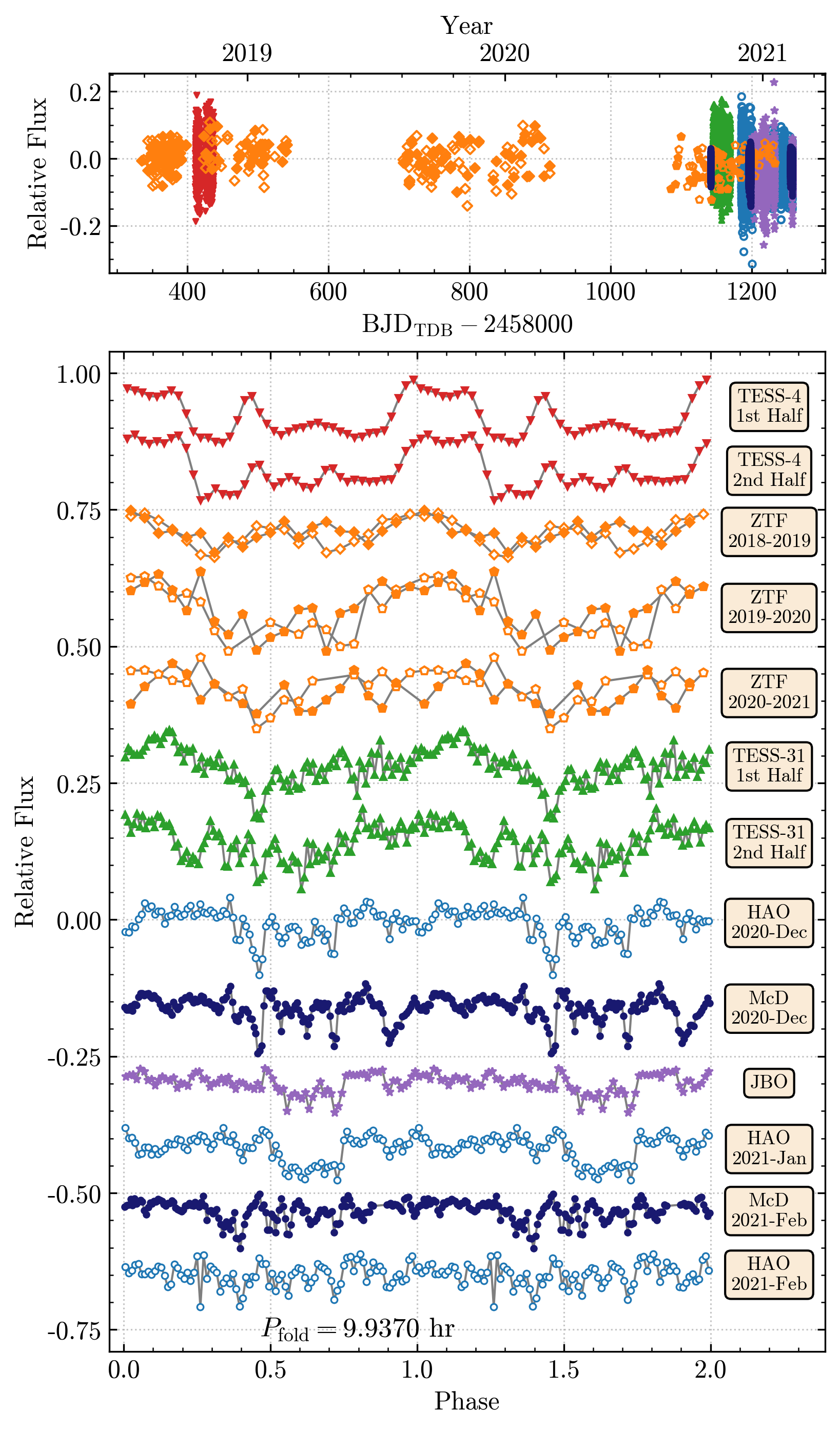}
	\caption{\textbf{Top}: Light curves from \tess (red and green triangles), ZTF (orange pentagons), HAO (blue open circles), JBO (purple stars), and McDonald Observatory (dark blue circles) showing the total time baseline covered by each data set. The \tess Sector 31 data shown in this panel are binned into 10-minute exposures to reduce scatter. \textbf{Bottom}: Light curves phase folded on the best-fit A-period, \Aperiod~hr, shown in rough chronological order from top to bottom, and labeled at right by their source. The \tess Sector 4 and 31 light curves are divided into first- and second-half segments to show changes in light curve structure over their durations, while the ZTF data are divided into three groups by observing season. The HAO and McDonald data are grouped by month. For the ZTF light curves, we display the $g$-band (filled symbols) and $r$-band (open symbols) separately. Each light curve has been vertically shifted for clarity, and two full phases have been plotted. \label{fig:allphot}}
\end{figure}

In Figure~\ref{fig:mcdphot} we present the detailed phased light curves obtained from McDonald and SAAO, and in Figure~\ref{fig:allphot}, we present the binned, phase-folded light curves obtained from all of our photometry sources. The light curves have been folded on the best-fit A-period, $P_{\mathrm{A}}=\Aperiod\,$hr, identified in Section~\ref{subsec:period} and marked with an asterisk in Table~\ref{tab:periods}. The phases are given with respect to an arbitrary date, $\mathrm{BJD_{TDB}}=2459141.504938$, chosen close to the start of the first McDonald run on 2020 October 19. We only generate binned, phase-folded light curves averaged over several days and nights for \tess, ZTF, HAO, and JBO due to having lower S/N or sparse sampling. \tess Sector 31 would be an ideal data set to observe cycle-to-cycle variations in dip structures due to its 2-minute cadence and near-continuous coverage over about 26 days, but \obj is too dim to observe individual dips within a single A or B-period cycle. For example, the largest variations observed in \obj have amplitudes of about 10\%, but the average point-to-point scatter among the \tess Sector 31 data points in about 18\%, so noise clearly dominates unless averaged out over several cycles.

McDonald light curves from the four consecutive nights in 2020 December and the four consecutive nights in 2021 February each span nearly 8 full cycles of $P_{\mathrm{A}}$, and show both broad and narrow dip-like features with depths of 5$-$10\% that line up in phase but also vary significantly in depth and shape between nights. This behavior is reminiscent of the dynamic orbit-to-orbit variations in dip structure seen in WD\,1145+017 \citep{Gaensicke2016,Rappaport2016,Gary2017}. The narrowest of the McDonald dips have durations of about 20 minutes, while the broadest features extend more than an hour. In contrast to the observed behavior of WD\,1145+017 which often shows intervals of inactivity between successive dip events, \obj appears to exhibit constant photometric variability when observed at both high speed and high signal-to-noise, likely the result of being occulted by dust clouds populating a wide range of phases within the orbit.

In Figure~\ref{fig:allphot}, the phase-folded and binned light curves from individual data sets, which span a total of about 2.5 years, also show variations in light curve structure over a broad range of timescales. We split each \tess light curve into first and second half components, each between 7.4 and 12.2 days long, to search for changes in light curve structure on these timescales. Between the first and second halves of \tess Sector 31, a large dip appeared at phase 0.2, which then disappeared again in the HAO observations which began just 16 days after \tess Sector 31 completed. At the same time, a sharp dip feature around phase 0.45 seems to persist from Sector 31 through our McDonald data (about 54 days total), only recently becoming less pronounced in JBO observations which began on 2020 December 16.

Since high-speed observations have revealed that dips can occur on timescales as short as 20 minutes, and changes in the modulation profile can occur on day to week-long timescales, the 30-min long exposure times of the \tess Sector 4 data will inevitably tend to smooth the light curve structure, while the $\approx$7-month long time baselines covered by each season of ZTF data will average out changes in the modulation profile. These effects are noticeable in their folded light curves which show considerably less detail, but still show general trends which are in agreement with the more resolved structures seen in the \tess Sector 31, McDonald, HAO, and JBO light curves. For example, the regions of lowest flux in \tess Sector 4 and ZTF are aligned in phase with the regions of highest dip activity in the other light curves. 

\begin{figure}[t!]
	\epsscale{1.15}
	\plotone{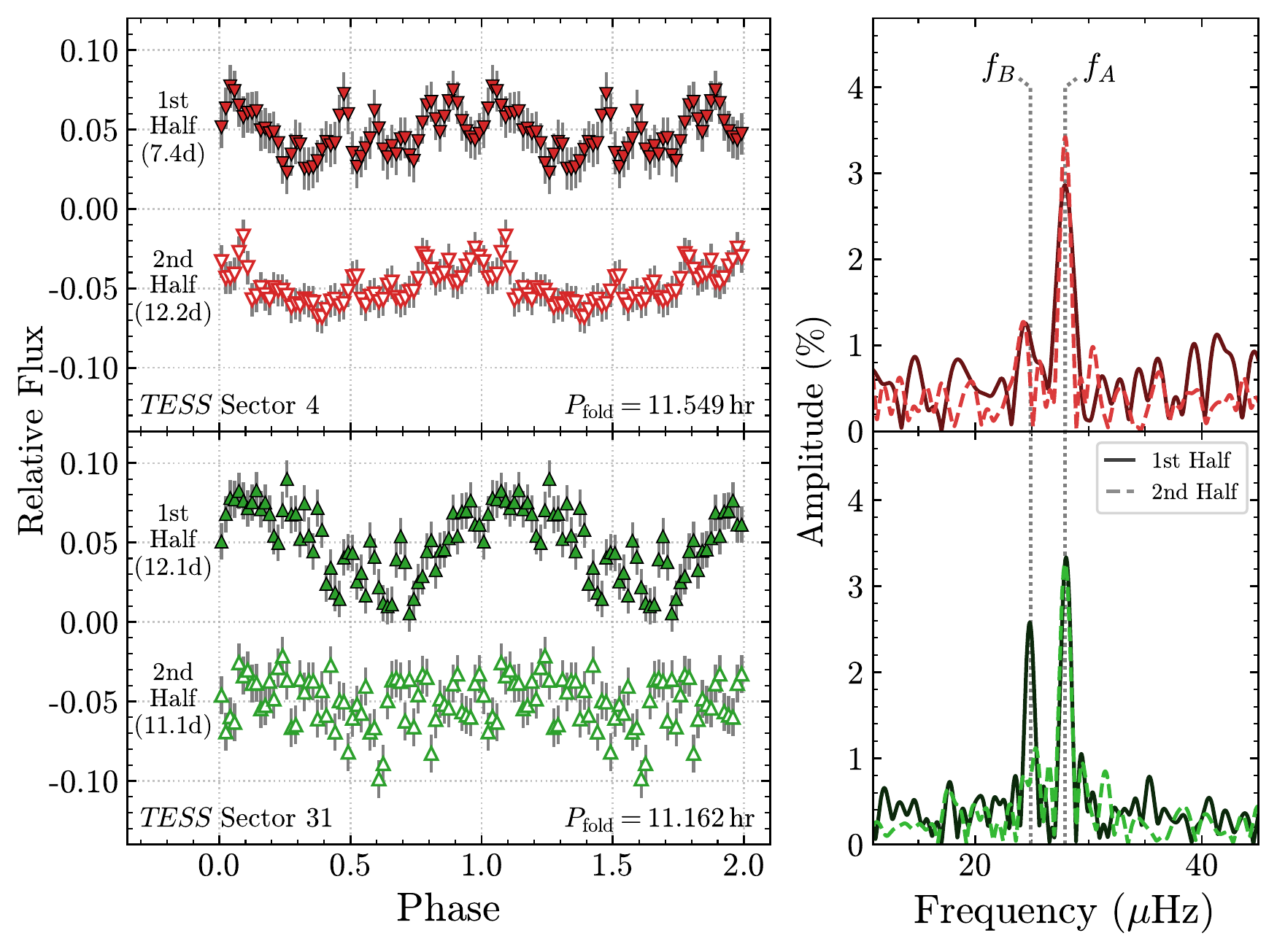}
	\caption{\textbf{Left}: The \tess Sector 4 (top) and Sector 31 (bottom) light curves phase-folded on the best-fit B-periods determined from each sector, and vertically shifted for clarity. We again split each \tess light curve into first- and second-half components to display changes that occur throughout their ${\approx}$26-day durations. The amount of data contained within each phase-folded segment is shown to the left of each light curve in parentheses. \textbf{Right}: The corresponding periodograms of each \tess sector, where solid lines indicate the first-half light curves, and dashed lines the second-half. Frequencies corresponding to the A and B periods ($f_A$ and $f_B$) are marked with vertical dotted lines. In \tess Sector 31, the B-period underwent a significant reduction in amplitude from 2.3$\pm$0.2\% in the first half to 0.8$\pm$0.2\% in the second half. \label{fig:Bfold}}
\end{figure}

Despite our extensive follow-up observations, we lack a conclusive $P_B$ detection in the periodograms of the MHJS photometry, nor do we readily identify any persistent features in our highest S/N light curves from McDonald and SAAO when folding on the B-period. This is possibly due to the relatively short durations of our individual observing runs, which fail to capture the full 11.2\,hr B-period in one night. Thus, if the 11.2\,hr period results from one or more localized dip features, these dips may not always be observed each night. On the other hand, if the 11.2\,hr period results from broad features, these may be diminished in amplitude by the routines we use to remove airmass trends in some of our nightly runs. Additionally, given the changes in period and harmonic structure of the $P_B$ peaks between the two \tess sectors, it is also possible that the activity level at this period changes with time, and our follow-up observations occurred during an interval of low activity at this period.

Since the \tess observations cover a large number of both $P_A$ and $P_B$ cycles without any daytime interruptions, they offer a better chance at averaging out variations related to the A-period and revealing structure when phase-folded on the B-period. In Figure~\ref{fig:Bfold}, we show the \tess light curves phase-folded on the best-fit $P_B$ values determined using {\sc Period04} for each sector (see Table~\ref{tab:periods}). We again divide each sector into first and second halves to reveal changes in structure over the ${\approx}$26-day duration of each sector, and also show the periodograms associated with each \tess light curve segment.

In both halves of the Sector 4 light curves, the B-period variations are only weakly detected, but they are clearly detected in the first half of Sector 31 where they appear nearly sinusoidal with an amplitude of 2.3$\pm$0.2\%. In the second half of Sector 31, however, the B-period amplitude decreased to 0.8$\pm$0.2\%, potentially revealing a rapid reduction in the activity level associated with this period just 16 days prior to the start of our HAO observations. It is possible the activity level remained low throughout our follow-up observations with HAO, McDonald, JBO, and SAAO, though this is difficult to confirm given the previously mentioned issues surrounding our ability to detect $P_B$ variability with ground based observations. Future ground-based observations with continuous coverage, such as with the Whole Earth Telescope \citep{Nather_1990}, would greatly aid in the detection and detailed characterization of the $P_B$ modulations.

\subsection{White Dwarf Atmospheric Properties} \label{subsec:atmos}

Previous estimates of the effective temperature and mass of \obj have been attempted using \gaia DR2 photometry and parallax while assuming a pure H or He atmosphere \citep{GF2019}. Since being identified as a metal-polluted He-atmosphere white dwarf whose dominant spectral features come from elements heavier than He (DZ), these estimates are no longer reliable for \obj. Using the same LDT spectrum presented in \citet{Guidry2021}, along with broadband optical and near-infrared photometry from \gaia eDR3 and PS1, we determine new atmospheric parameters for \obj.

\begin{deluxetable}{lcc}
\tablenum{6}
\tablecaption{Atmospheric Properties of \obj}
\label{tab:atmos_params}
\tablewidth{0pt}
\tablehead{\colhead{Parameter} & \colhead{Value} & \colhead{Uncertainty}} 
\startdata
$T_{\mathrm{eff}}$ (K)      & 7630        & 140   \\
$\log(g)$ (cgs)             & 8.245       & 0.035 \\
$M_{\star}$ ($M_{\odot}$)   & 0.731       & 0.023 \\
$R_{\star}$ ($R_{\odot}$)   & 0.0107      & 0.0002\\
\calcium                    & $-$9.55     & 0.12  \\
\hydro                      & ${<}{-}$3.5 & --    \\
$\tau_\mathrm{cool}$ (Gyr)  & 1.84        & 0.17  \\
\enddata
\end{deluxetable} 

\begin{figure}[t!]
	\epsscale{1.1}
	\plotone{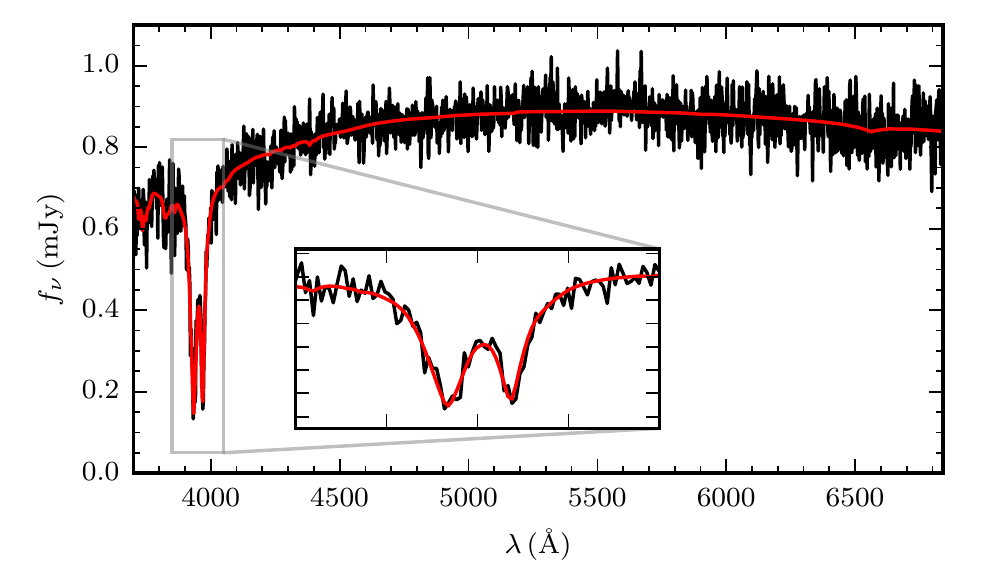}
	\caption{The LDT spectrum (black) along with the best-fit model (red), with an inset zoomed in on the Ca\,H and K features.  \label{fig:specfit}}
\end{figure}

\begin{figure}[t!]
	\epsscale{1.15}
	\plotone{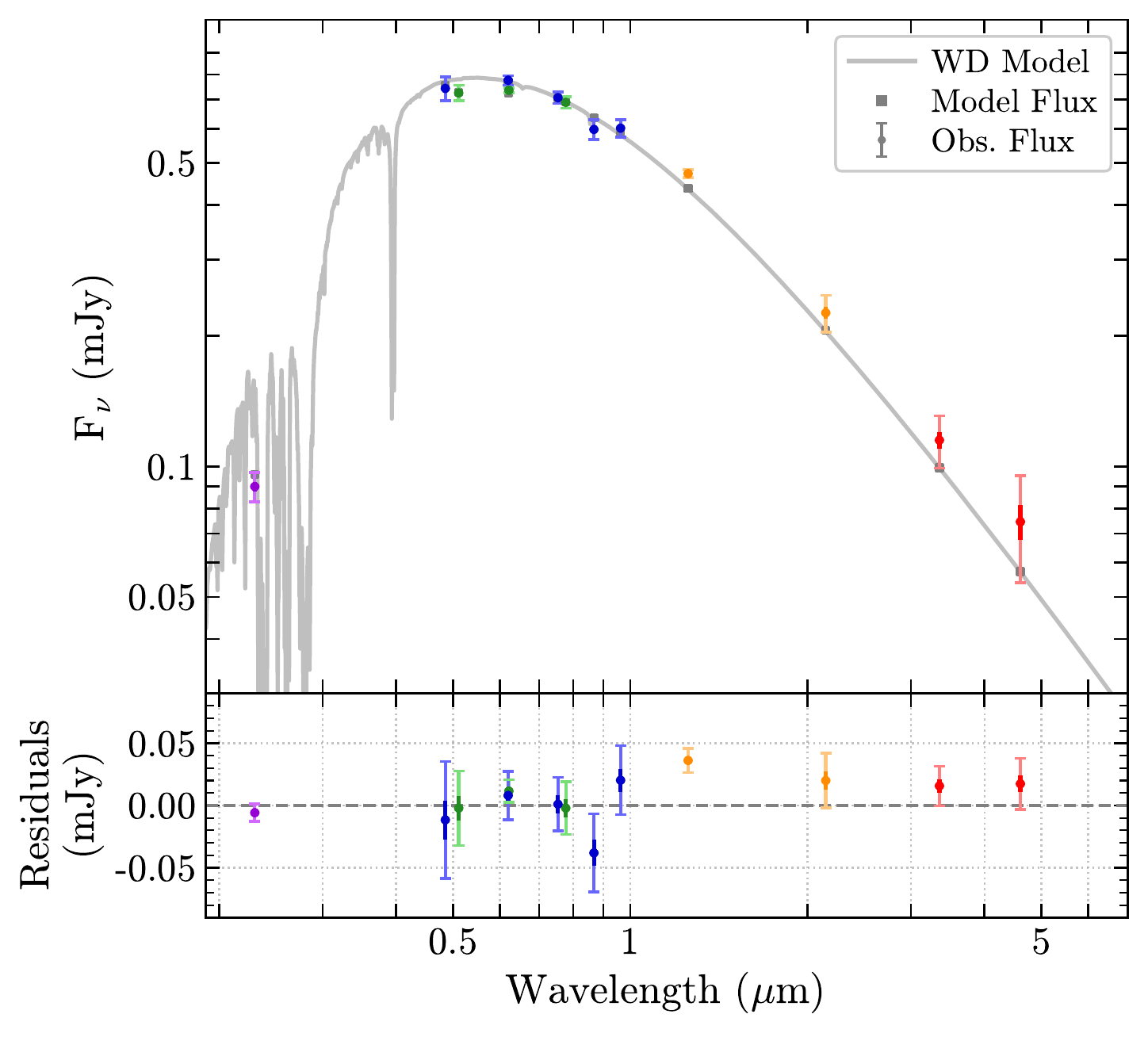}
	\caption{\textbf{Top}: The observed SED for ZTF\,J0328$-$1219 including {\em GALEX} (purple), Pan-STARRS1 (blue), \gaia eDR3 (green), VHS (orange), and unWISE (red) photometry. The best-fit white dwarf model is plotted in grey, with calculated photometric fluxes in each passband shown with dark grey squares. The dark and light-colored error bars show the 1-sigma and 3-sigma uncertainties, respectively, for each measurement. We only used the PS1 and \gaia photometry to constrain the white dwarf atmospheric fit, which produces a slightly poorer fit to the VHS and {\em WISE} photometry. \textbf{Bottom}: Residuals between the observed and model flux values in each passband. We find no evidence for an infrared excess in the {\em WISE} $W1$ and $W2$ bands within their 3-sigma uncertainty limits, though a small but significant excess is present in the VHS $J$-band. \label{fig:sed}}
\end{figure}

We follow the fitting procedure outlined in \citet{Coutu2019} and employ the DZ model atmospheres of \citet{Blouin2018}. We first derive an effective temperature, \teff, and solid angle, $\pi R^2 / D^2$, using the \gaia eDR3 and PS1 photometry. From this solid angle and the \gaia parallax, we obtain the white dwarf radius. We then evaluate the corresponding white dwarf mass and \logg using theoretical white dwarf structure models \citep{Fontaine2001}, assuming a C/O core, a helium envelope of $\log (M_{\rm He}/ M_{\star}) = -2$, and a thin hydrogen envelope of $\log (M_{\rm H}/ M_{\star}) = -10$. Keeping \teff and \logg fixed to those values, we then adjust the calcium abundance, \calcium~$=\log(n_{\mathrm{Ca}}/n_\mathrm{He})$ where $n$ is the number density, to the LDT spectrum. H is included in our model atmospheres assuming an abundance \hydro~$=\log(n_{\mathrm{H}}/n_\mathrm{He})$ that corresponds to the visibility limit of the H$\alpha$ line \citep{Coutu2019}, which is not detected in the LDT spectrum. In addition, all elements from C to Cu are included assuming chondritic abundance ratios with respect to Ca. An iterative procedure which alternates between these two steps (the photometric fit and the adjustment of the abundances to the spectrum) is then performed to converge on a best-fit solution. The final values are summarized in Table~\ref{tab:atmos_params}, while Figure~\ref{fig:specfit} shows the DZ model fit to the LDT spectrum, while Figure~\ref{fig:sed} shows the DZ model fit the to SED.

For \obj we derive $T_{\mathrm{eff}}=7630\pm140\,$K,  $M_{\star}=0.731\pm0.023\,M_{\odot}$, and $R_{\star}=0.0107\pm0.0002\,R_{\odot}=1.17\pm0.02\,R_{\earth}$. Due to the inclusion of atmospheric metals and H, this is significantly cooler and less massive than the pure He-atmosphere parameters derived by \citet{GF2019} using \gaia DR2 photometry and parallax ($T_{\mathrm{eff}}=8551 \pm 161\,$K and $M_{\star}=0.855\pm0.029\,M_{\odot}$). This trend is in line with previous results \citep[e.g.,][]{Coutu2019} and is due to the increased He$^-$ free--free opacity resulting from the additional free electrons provided by the metals and H. We chose not to use the VHS and unWISE photometry in our fitting process so we could better assess whether an infrared excess is present. If we include these data in the fit, however, we find a 2-sigma drop in $T_{\mathrm{eff}}$ by about 300\,K, producing a better fit to the VHS and unWISE photometry while slightly degrading the fit to the {\em Gaia} and PS1 photometry.

We estimate a Ca abundance of \calcium~$=-9.55\pm0.12$, consistent with abundances of known metal-polluted He-atmosphere white dwarfs in this temperature range \citep{Blouin2020,Coutu2019,Hollands2017,Dufour2007}, and put an upper limit on the H abundance of \hydro~$<-3.5$ based on the non-detection of Balmer lines in the LDT spectrum. Assuming the white dwarf has followed a standard single-star evolutionary pathway and has a thin H envelope ($\log M_{\rm H}/M_{\star}=-10$), we find a white dwarf cooling age of $\tau_{\mathrm{cool}}=1.84\pm0.17\,$Gyr using the evolution models described in \citet{Blouin2020b}. 

Using the He convection zone masses ($M_{\mathrm{CVZ}}$) and Ca diffusion timescales ($\tau_{\mathrm{Ca}}$) for helium-atmosphere white dwarfs in \citet{Koester2009}, we find $\log(M_{\mathrm{CVZ}}/M_{\star})=-4.75$ and $\log(\tau_{\mathrm{Ca}}[\mathrm{yr}])=6.46$ at the effective temperature of \obj. Assuming heavy metal abundance ratios outlined in \citet{Zuckerman2010}, this implies that $4.4\times10^{21}\,$g of heavy metals currently reside within the outer He convection zone, or about 0.5\% the mass of Ceres. 

Steady-state accretion episodes around white dwarfs are expected to last up to $10^4-10^6\,$yr \citep{Girven2012}, and often times are expected to be much shorter \citep{Veras_2020_2}, making it highly unlikely that the current \calcium is being sustained with steady-state accretion. More likely, the \calcium represents some combination of residual Ca from previous accretion episodes, and new Ca being presently accreted. Steady-state accretion would imply a total mass accretion rate of $\log(\mathrm{\dot{M}[{\rm g\,s}^{-1}]})=7.7$, consistent with accretion rates for white dwarfs with similar temperatures and atmospheric compositions \citep{Dufour2007,Farihi_2016_1,Hollands2017,Xu2019}, but the current accretion rate might actually be orders of magnitude higher or lower. Compared to the steady-state accretion rate of WD\,1145+017 at $\log(\mathrm{\dot{M}[{\rm g\,s}^{-1}]})=10.6$ \citep{Xu_2016_1} whose diffusion timescales are of order $10^5$\,yr, however, \obj is much lower. This might be the result of less overall accretion in \obj due to having lower-mass bodies in orbit, or perhaps due to the bodies orbiting further away from the white dwarf and undergoing less disruption, but it is also possible that past accretion episodes have contributed to the currently observed abundances in these objects given their long diffusion timescales.

We note that even though the DZ model provides a good fit to the observed Ca\,H and K lines, some caution must still be taken when interpreting the calculated Ca abundance. WD\,1145+017 exhibits prominent circumstellar absorption components \citep{Xu_2016_1} which evolve significantly over time \citep{Redfield_2017_1,Cauley_2018}. We do detect some narrow atomic absorption features consistent with circumstellar absorption in our high-resolution MIKE spectrum (see more detailed analysis in Section~\ref{subsec:circumstellar}), the strongest of which have equivalent widths of about 0.2\AA. This is small compared to the ${\approx}45$\AA\ combined equivalent width of the photospheric Ca~H and K lines, so we suspect the impact of circumstellar absorption features on our measured \calcium is negligible.

We also note a weak H$\alpha$ feature is present in our high-resolution MIKE spectrum (see Section~\ref{subsec:circumstellar} and Figure~\ref{fig:soarspec}) and we attempt to include it in our atmospheric modeling of the white dwarf for better constraints on \hydro, but the narrow line is poorly fit with current atmospheric models. To match the depth of the narrow H$\alpha$ feature, our models suggest that a broad H$\alpha$ component should also be present, which is not observed. Such mismatches have been seen before in metal polluted white dwarfs, such as the narrow He-I lines observed in GD\,362 \citep{Zuckerman_2007,Tremblay_2010} and in GD\,16 and PG\,1225-079 \citep{Klein_2011}. Therefore, we do not consider the mismatch a sign of a non-atmospheric origin for H$\alpha$, but rather a difficulty with atmospheric modeling. We still believe that our calculated upper limit for \hydro of $-$3.5 at the photosphere is valid since it is based on the lack of a broad H$\alpha$ absorption feature in the low-resolution LDT spectrum.

\subsection{Circumstellar Gas and Dust} \label{subsec:circumstellar}

In addition to atmospheric metal pollution, other observational signatures of planetary debris around white dwarfs include infrared excesses due to the presence of warm dust disks \citep[e.g.][]{Zuckerman1987,Dennihy_2020_1}, metallic emission lines from hot gaseous components of the dust disks \citep[e.g.][]{Gaensicke2006,Melis2020}, and metallic absorption lines from circumstellar gas \citep[e.g.][]{Debes_2012_2,Xu_2016_1}. WD\,1145+017, the prototypical transiting debris white dwarf, exhibits an infrared excess \citep{Vanderburg2015} and circumstellar gas absorption \citep{Xu_2016_1}, but has not been observed to exhibit metallic emission lines. 

\begin{figure}[t!]
	\epsscale{1.15}
	\plotone{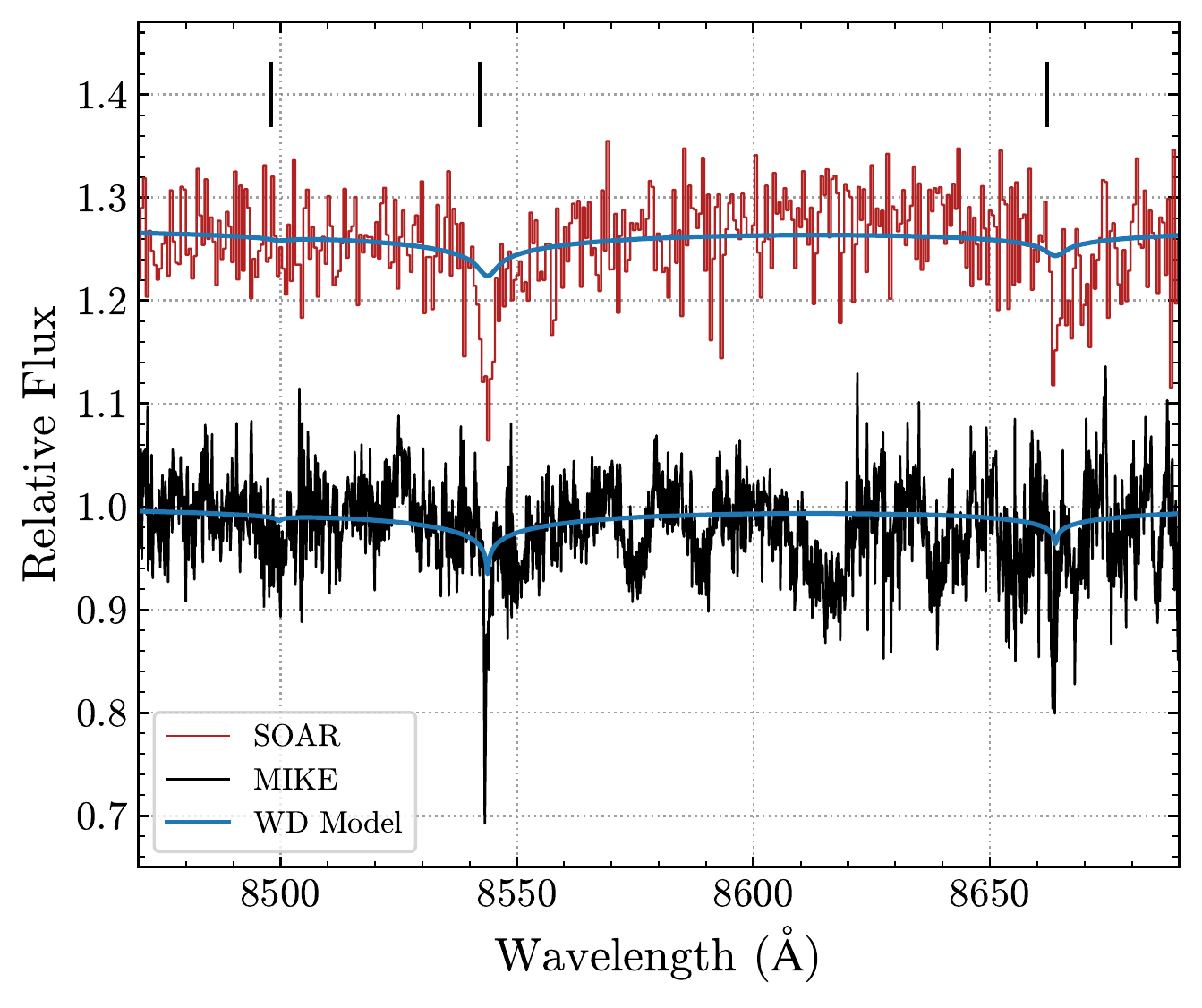}
	\caption{The SOAR (top) and MIKE (bottom) continuum-normalized spectra centered on the Ca infrared triplet region, and vertically shifted for clarity. The best-fit white dwarf model spectrum convolved with a Gaussian kernel to the respective resolutions of SOAR and MIKE is overplotted in blue and velocity-shifted to match the measured velocity of the H$\alpha$ line (58.8\,km~s$^{-1}$). Vertical black lines denote the rest-wavelengths of each Ca infrared triplet component. CCD fringing is prevalent in the MIKE spectrum at these wavelengths, producing many of the broad features not observed in the SOAR spectrum, though detections of absorption at 8542\AA, and possibly at 8662\AA, are seen in both. We find no evidence for Ca triplet emission in these observations. \label{fig:triplet}}
\end{figure}

We find no evidence for gaseous emission in any of our new spectroscopic observations from MIKE or SOAR (see Figure~\ref{fig:triplet}). Also, in the \wise $W1$ and $W2$ bands, the observed flux agrees within 3-sigma with the best-fit DZ model (see Figure~\ref{fig:sed}). Using unWISE $W1$ and $W2$ photometry, \citet{Xu2020} also do not find a significant magnitude or color excess for \obj, although they assumed a pure hydrogen-atmosphere white dwarf model. We do find a small but significant excess of 0.04\,mJy (11.1-sigma) in the VHS $J$-band. Larger $J$-band flux excesses relative to $W1$ and $W2$ excesses are sometimes seen for white dwarfs with cool sub-stellar companions \cite[see Figure 1 of][]{Xu2020}, but this may also be the result of underestimated $J$-band uncertainties coupled with a white dwarf atmospheric fit that is slightly too hot. If we include the VHS and unWISE photometry in the atmospheric fitting process described in Section~\ref{subsec:atmos}, the best-fit white dwarf model becomes about 300K cooler, and the VHS $J$-band then agrees within 4.3-sigma. To further investigate whether an infrared excess is present in \obj, higher-S/N observations in the near and mid-infrared, such as with {\em JWST}, will be needed.

\begin{figure}[t!]
	\epsscale{1.15}
	\plotone{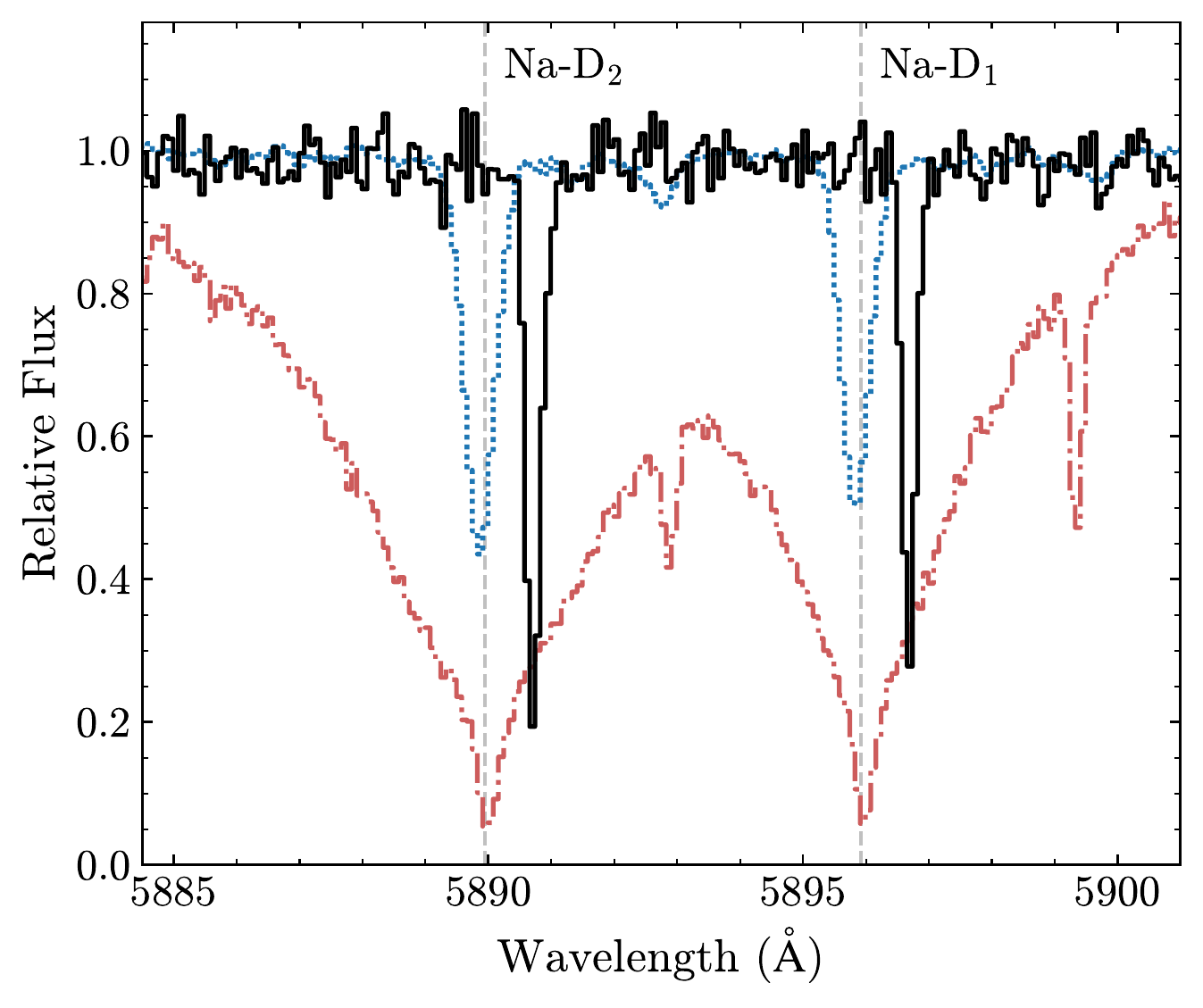}
	\caption{The continuum-normalized MIKE spectra centered on the Na~D doublet for \obj (solid black line) and the two nearest comparison stars, HD\,21634 (dotted blue line) and Gaia\,J0328$-$1216 (dash-dot red line). The vertical dashed lines denote the rest wavelengths of the Na D doublet. Neither comparison star shows any Na~D absorption --- in addition to their photospheric components --- at the same velocity as those in \obj, ruling out an interstellar origin. \label{fig:sodium}}
\end{figure}

We detect prominent and narrow Na-I D lines in the MIKE spectrum of \obj, with a full-width-half-maximum (FWHM) of $0.24$\AA\ and equivalent widths of $0.18\pm0.01\,$\AA\ and $0.24\pm0.01\,$\AA\ for D$_1$ (5895\AA) and D$_2$ (5889\AA), respectively (see Figure~\ref{fig:sodium}). To determine whether these lines are of interstellar or circumstellar origin, we compare their line shapes and velocities to the Na D lines observed in our nearby comparison stars. Using the 100\,$\mu$m IRAS dust maps \citep{Schlegel_1998} and extinction values calculated by \citet{Schlafly2011}, we find that the line-of-sight Galactic extinction at the coordinates of \obj is $A_V=0.221$. At a distance of just 43\,pc, however, it is unlikely that a significant amount of ISM material lies between us and \obj, and the 3D dust maps of \citep{Green2019} predict no extinction at the coordinates and distance of \obj. The two comparison stars nearest to \obj on sky, Gaia\,J0328$-$1216 and HD\,21634, are the most suitable for comparing Na line shapes and velocities since they would likely sample the same ISM material and have the most similar line-of-sight Galactic extinctions compared to \obj with $A_V=0.234$ and $A_V=0.253$, respectively. Both comparison stars are about three times further away than \obj so they might experience more ISM absorption, but the 3D dust maps of \citet{Green2019} again predict no extinction at the coordinates and distances of these objects. The same is true for the two comparison stars farthest from \obj on sky, HD\,22243 and HD\,21917, but which have lower line-of-sight Galactic extinctions ($A_V$ of 0.091 and 0.156, respectively).

To establish our comparison stars as reliable radial velocity (RV) references, we first compared their measured velocities to catalogued \gaia DR2 values. One RV standard, GJ\,908 (spectral type M1 at $-$71.13 km~s$^{-1}$), was observed on 2020 December 1. We measured the velocities of our comparison stars from 22 orders of our blue spectra by cross-correlation with GJ\,908. Our measured radial velocities are close to those in DR2: Gaia\,J0328$-$1216: $1.08 \pm 0.14$ versus $2.3 \pm 0.93$ km~s$^{-1}$, HD\,21634: $-6.35 \pm 0.42$ versus $-7.37 \pm 0.49$ km~s$^{-1}$, and HD\,21917: $-58.0 \pm 0.2$ versus $-57.9 \pm 0.18$ km~s$^{-1}$. We exclude HD\,22243 from our RV analysis due to its exceptionally broad lines, though we still use it to search for narrow Na features indicative of ISM absorption. We also measured the velocities of several individual lines in our comparison stars by fitting them with Voigt profiles. We did this for the Na-I~D doublet, the Ca-I 6717\AA\ line, and H$\alpha$, and found velocities in each case that are consistent with the cross-correlation velocities, confirming photospheric origins for these lines. Hereafter, we use our measured, cross-correlation RVs to compare with \obj.

\begin{deluxetable}{lccc}
	\tablenum{7}
	\tablecaption{Radial Velocities of \obj from MIKE}
	\label{tab:rv}
	\tablewidth{0pt}
	\tablehead{\colhead{Line} & \colhead{$\lambda_{\mathrm{rest}}$} & \colhead{RV} & \colhead{Origin} \\
	\colhead{} & \colhead{(\AA)} & \colhead{(km~s$^{-1}$)} & \colhead{} }
	\startdata
	H$\alpha$          & 6562       & 58.8$\pm$0.8 & atmospheric \\
	Na-I               & 5889, 5895 & 37.4$\pm$0.5 & circumstellar \\
	Ca-II$^{\dagger}$  & 8542       & 40.1$\pm$1.5 & circumstellar \\
	Ca-II$^{\ddagger}$ & 3933, 3968 & 51.5$\pm$3.5 & blended? \\
	\enddata
	\tablenotetext{\dagger}{Velocity shown is for the narrow line component.}
    \tablenotetext{\ddagger}{Velocity shown is for the narrow H and K line cores.}
\end{deluxetable}

\begin{figure}[t!]
	\epsscale{1.1}
	\plotone{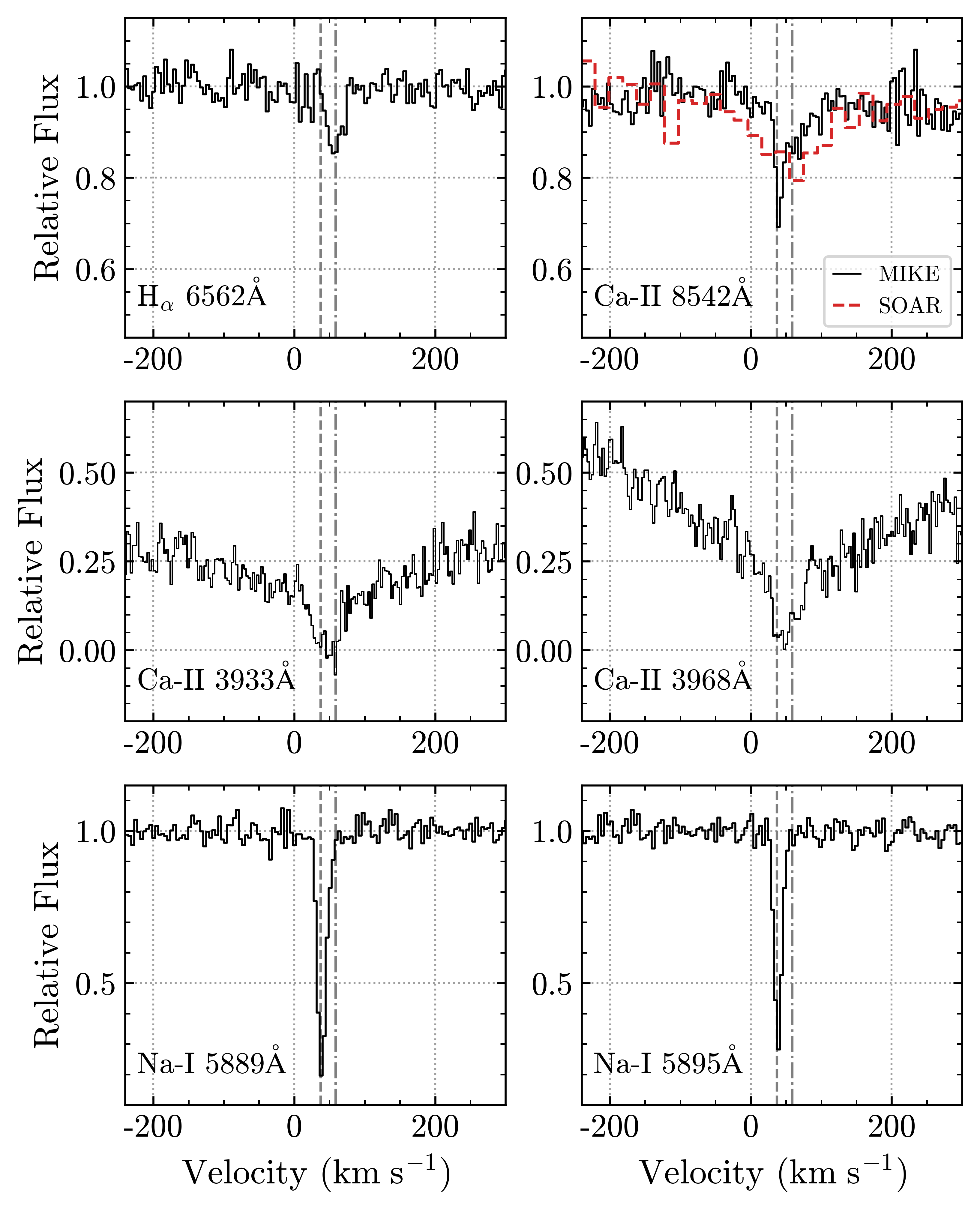}
	\caption{The continuum-normalized MIKE spectrum (solid black) of \obj near various lines with measured velocities (see Table~\ref{tab:rv}). For the 8542\AA\ Ca triplet component, we also show the continuum-normalized, lower-resolution SOAR spectrum (dashed red) which detects the same feature. The 3933 and 3968\AA\ panels (Ca-II K and H respectively) highlight the narrow line components, but also exhibit a small portion of the broad components of these lines which cause the upward slopes away from line center. Vertical lines denote the measured velocities of the Na and H$\alpha$ features from MIKE spectra, with the dashed line showing the 37.4 km~s$^{-1}$ velocity of the Na lines, and the dash-dotted line showing the 58.8 km~s$^{-1}$ velocity of the H$\alpha$ feature. \label{fig:soarspec}}
\end{figure}

We measured the absolute radial velocity of the Na~D lines in \obj by cross-correlation against the stellar Na lines of four comparison stars: HD\,21917, HD\,21634, Gaia\,J0328$-$1216, and GJ\,908. We used only the wavelength region between 5886 and 5899\,\AA\ for cross correlation, and we do not perform any corrections to the RVs for the gravitational redshift caused by the white dwarf. The average absolute radial velocity of the Na~D lines in \obj is $37.4\pm0.5$ km~s$^{-1}$. 

We also detect a weak H$\alpha$ line in \obj with an equivalent width of only $0.08\pm0.01\,$\AA, but a FWHM of 0.6\,\AA\ (see Figure~\ref{fig:soarspec}). We measure its velocity to be $58.8\pm0.8$~km~s$^{-1}$, a difference of $21.4\pm1.0$~km~s$^{-1}$ with respect to the Na~D features in the MIKE spectrum. In the blue-side MIKE spectrum, we also detect Ca\,H and K lines with broad wings and narrow components at their cores, consistent with the LDT observations. The S/N in the Ca\,H and K order is not good enough, however, to measure the velocities precisely, and cross-correlation with the reference stars does not work well because of the vastly different shapes of the lines. Still, we attempt to measure the H and K velocity by fitting a Voigt profile to the line core and obtain a velocity of $51.5\pm3.5$ km~s$^{-1}$. We caution that in addition to the noisy detection, the cores of these lines may be blended with circumstellar absorption features if present, further complicating this velocity measurement.

Two of the Ca infrared triplet lines, 8542 and 8662\AA, are also detected in the MIKE spectrum. The spectrum is considerably noisier at these wavelengths and significant CCD fringing is present, but we detect similar features in our lower-resolution SOAR spectrum, which has only 0.3\% peak-to-peak fringing at these wavelengths, leading us to believe these features are real. The stronger of the two lines at 8542\AA\ has a narrow component with FWHM ${\sim}0.25$\AA, similar to that of the Na~D lines, superimposed on a broader, slightly redder component (see Figure~\ref{fig:soarspec}). By simultaneously fitting the narrow and broad line components with two Voigt profiles of different widths, and comparing them to the position of the stellar lines in the comparison stars, we measure the velocity of the narrow and broad components of the Ca~8542\AA\ line as $40.1\pm1.5$  and $59.5\pm3.5$ km~s$^{-1}$, respectively. In our SOAR spectrum, the 8542\AA\ feature is similar in strength to the MIKE feature, though the narrow component is not detected at the lower resolution (see Figure~\ref{fig:soarspec}). We once again fit a Voigt profile to the 8542\AA\ component in the SOAR spectrum, and measure a velocity of $59.6\pm8.9$ km~s$^{-1}$, in agreement with the broad component observed by MIKE.

The Na~D lines provide the most compelling evidence for a detection of circumstellar gas absorption. They are much stronger than predicted by our DZ model, and much narrower than would be expected if photospheric. We can also rule out interstellar absorption since none of the comparison stars show similarly narrow Na~D absorption --- in addition to their photospheric absorption --- at the same velocity as \obj (see Figure~\ref{fig:sodium}). This includes the two nearest comparison stars, HD\,21634 and Gaia\,J0328-1216, which are most likely to sample the same ISM material as \obj, and also the two comparison stars further away, HD\,22243 and HD\,21917.

The H$\alpha$ feature is likely to be in the white dwarf atmosphere given its velocity is 21\,km~s$^{-1}$ red-shifted with respect to the Na~D and calcium triplet lines. This difference is smaller than expected since the gravitational redshift for a 0.73\,$M_{\odot}$ white dwarf would be about 44\,km~s$^{-1}$, but this also assumes that the Na~D lines come from circumstellar gas traveling perpendicular to our line of sight towards the white dwarf, which may not be the case if the gas has an eccentric orbital configuration. Smaller velocity differences may also be observed if the gas is located sufficiently deep in the gravitational well of the white dwarf, but this effect has only been observed for high ionization species \citep{Gaensicke2012}, and likely does not apply here. Long-term spectroscopic monitoring and modeling of the Na~D line velocities, such as the studies that have been done for WD\,1145+017 \citep{Redfield_2017_1,Cauley_2018,Fortin-Archambault_2020} and WD\,1124-293 \citep{Debes_2012_2,Steele2020}, can be used to probe the geometry and column densities of the circumstellar gas in \obj. 

The narrow component of the Ca-II 8542\AA\ feature is likely of circumstellar origin as well, since its velocity is most in agreement with the Na~D lines and has a similar FWHM, but the broader component detected by both MIKE and SOAR is more in agreement with the H$\alpha$ velocity. It is currently unclear if that suggests some of the Ca infrared triplet absorption has an atmospheric origin, since the DZ model predicts much weaker absorption than what is observed. It is possible that the circumstellar gas exhibits an asymmetric Doppler-broadened line profile that can account for both of the 8542\AA\ line components \citep[e.g.][]{Xu_2016_1}, but it would be puzzling why a similar velocity distribution is not observed for the much stronger Na~D lines. With continued spectroscopic monitoring at high resolution, it will be interesting to see if any velocity variations in the Na~D lines also occur to one or both components of the Ca-II 8542\AA\ line.

\section{Discussion} \label{sec:discussion}

\subsection{Orbiting Debris} \label{subsec:debris}

The evidence for planetary debris in orbit around \obj consists of atmospheric metal pollution, periodic optical variability at 9.93 and 11.2\,hr, and the presence of circumstellar gas. As seen in Figures~\ref{fig:mcdphot} and \ref{fig:allphot}, the characteristics of the optical variability closely resemble those in WD\,1145+017 \citep[see][]{Gaensicke2016,Rappaport2016,Rappaport2018}, where small variations in the detailed transit shapes are likely occurring on orbit-to-orbit timescales, while more dramatic changes can occur on week, month, and year-long timescales. In spite of these similarities, \obj also exhibits unique characteristics which bring into question the source of its variability. Notably, \obj appears almost continuously variable (see Figure~\ref{fig:mcdphot}), whereas WD\,1145+017 in most cases exhibits intervals of quiescence between successive dip events. This may suggest that \obj is continuously being occulted by clouds of planetary debris which populate a wide range of orbital phases, but here we first consider alternate explanations before exploring the planetary debris scenario further.

\subsection{Other Possible Explanations for Variability}

\subsubsection{Rotational Modulation} \label{subsec:rotation}

White dwarfs are known to exhibit periodic photometric variability due to stellar rotation combined with magnetism or some form of surface inhomogeneity \citep[see e.g.][]{Maoz2015,Kilic2015,Hermes_2017_2,Reding2020,Gaensicke2020}. Typical white dwarf rotation periods range from 0.5--2.0\,days \citep{Hermes_2017_1}, so both the 9.93\, and 11.2\,hr periods are feasible options for a rotation period in \obj. Already, however, the presence of two periods eliminates rotational modulation as the sole source of variability, since it can only account for one of the observed periods. Additionally, the vast majority of known and suspected rotationally modulated white dwarfs exhibit persistent and near-sinusoidal variations in their light curves, but \obj exhibits both highly non-sinusoidal variations and rapid evolution in light curve structure on timescales of 1-day or less.

Due to the complexity of the light curve structure, however, we have yet to identify individual dip-like features in our ground-based followup observations associated with the 11.2\,hr B-period. The \tess light curves show a significant detection at this period, but are too noisy to analyze individual B-period cycles in detail. The phase-folded light curves show some details of the B-period modulation profile, which appears to change significantly in both shape and amplitude between \tess sectors and between the first and second halves of individual sectors (see Figure~\ref{fig:Bfold}), behavior which is not expected for rotationally modulated white dwarfs. These changes are also evidenced by the presence of harmonics in the \tess Sector 4 periodogram, indicating sharper features in the light curve, which are not seen in the \tess Sector 31 periodogram. Lastly, the broader-than-expected peaks seen in the periodogram (see Section~\ref{subsec:period}) also suggest changes to the modulation profile, changes in the period, or the presence of multiple closely spaced periods, all of which rule against rotational modulation. Therefore, we strongly suspect the 11.2\,hr period to most likely be associated with transiting dust-emitting debris, but we still consider rotational modulation a possibility for $P_B$ until more detailed observations are carried out which can better characterize the 11.2\,hr modulations.

\subsubsection{Cataclysmic Variable} \label{subsec:CV}

The near-constant photometric variability in \obj looks qualitatively similar to the photometric variability observed in some cataclysmic variables (CV). Unlike CVs, however, \obj does not exhibit any spectral emission lines that would indicate mass accretion from a companion stellar or sub-stellar object \citep[e.g.][]{Szkody2009}, or emission from the chromosphere of an irradiated companion \citep[e.g.][]{Longstaff2017, Longstaff2019}. 

In addition, the presence of a sub-stellar companion is strongly disfavored due to the lack of a significant infrared excess (Figure~\ref{fig:sed}). We generated composite SEDs for a white dwarf and L/T-dwarfs using the best-fit DZ model from Section~\ref{subsec:atmos} and L and T-dwarf templates from the SpeX Prism library \citep{Burgasser2014}. We combined the spectra following the procedure described in \citet{Casewell2017}, normalizing the white dwarf model to absolute $g$-band magnitudes from \citet{Holberg2006}, and the L and T-dwarf templates to absolute $J$-band magnitudes from \citet{Dupuy_2012_1}. 

The SpeX templates extend out to just 2.6\,$\mu$m, so we exclude the {\em WISE} photometry from our comparison with the composite SEDs. Within the 3-sigma uncertainty limits of the $J$-band and $K_{\mathrm{s}}$-band photometry, we can exclude the presence of a companion with spectral type earlier than T7. We have not carried out any time series spectroscopy to look for radial velocity variations, but we note that the radial velocity of the Ca~8542\AA\ line from the SOAR spectrum and the broad component of the Ca~8542\AA\ line from the MIKE spectrum are consistent with no change in velocity.

\subsection{Source of the Dusty Effluents}

Ultimately, we find that the periodic variability at both the 9.93 and 11.2\,hr periods is best explained by planetary debris transiting the white dwarf. We further postulate that the photometric variability itself is caused by dust, either entrained in the orbiting debris or emitted continuously by one or more discrete bodies. This general scenario is supported by the presence of atmospheric metal pollution, the highly non-sinusoidal and volatile nature of the periodic photometric variability, and the detection of circumstellar gas. 

The orbital semi-major axes corresponding to periods of 9.937 and 11.162\,hr are $2.11\,R_{\odot}$ and $2.28\,R_{\odot}$, respectively\footnote{If we assume that the debris is being fed by the disruption of two planetesimals orbiting at $2.11\,R_{\odot}$ and $2.28\,R_{\odot}$ from the host star, respectively, we can set a rough upper limit on the masses of the these bodies if we require that the two orbits be separated by more than $\sim$10 Hill's sphere radii of the bodies. This is a minimum requirement for long-term dynamical stability. From this, we can conclude that the masses of these bodies must be less than Jupiter's mass, which is not interestingly restrictive.}. At these distances, and in the case where no shielding of the white dwarf flux by intervening dust or gas occurs, the equilibrium temperatures of absorbing bodies could range from 385\,K to 700\,K for orbital periods near 9.93\,hr, and somewhat lower for orbiting bodies with periods near 11.2\,hr. These temperatures are summarized in Table~\ref{tab:temps}, and raise the issue of what kinds of materials could survive and continue to emit dusty effluents in this environment. We have also included in Table~\ref{tab:temps} the \citet{Chiang1997} formalism for equilibrium temperatures within a flat, optically thick disk that can partially shield an orbiting body from direct radiation coming from the host star. In order for this to be applicable, there must be a dusty disk, and the body should be physically smaller than the disk height. We currently have no direct evidence for the former and little information about the latter.

In order to guide our thinking about what types of bodies might emit dust at these temperatures, we consider the properties, including composition, of comets and asteroids.  In Table~\ref{tab:aster_comets} we present a broad-brush and simplified, but hopefully useful, comparison between these two types of bodies that are seen in the Solar System.  Comets are mostly fragile material whose main volatiles are H$_2$O, CO, and CO$_2$ ices.  The sublimation of these bodies occurs well below the environmental temperature of \obj, and hence would not last very long before completely evaporating.  By contrast, many of the minerals expected to be found in asteroids, some examples of which are given in Table~\ref{tab:aster_comets}, have sublimation temperatures above about 1200 K.  These temperatures are too high to expect substantial emission of heavy molecules that might condense into dust in $\sim$10-hr obits around \obj.

We know of no obvious candidate materials with sublimation temperatures in the vicinity of $400-600$~K. Solutions to this apparent paradox include the following possibilities.  (1) If the material responsible for the dust is in a ring of debris, then collisions can both eject pulverized fine particles into the surrounding space, or expose pristine volatiles lying beneath the surface of the body which had previously shielded this material from the stellar radiation.  (2) Volatile materials beneath the surface layers, and therefore largely shielded from the stellar radiation, may nonetheless be slowly heated to the point where the pressure buildup blows away and ejects the overlying material. (3) In a related scenario, if orbit migration of a comet is rapid compared to the timescale for heat to penetrate to the comet center, an exotic model invoking deep pits can be invoked to preserve volatile-based dust release at the present time. (4) If the dust-emitting body is indeed orbiting completely inside a dusty disk, then the equilibrium temperature of the body given by the \citet{Chiang1997} expression may be sufficiently low that dust-ladened ices of CO$_2$ and H$_2$O may survive the radiation of the white dwarf. Lastly, (5) if a body is nearly filling its critical potential lobe, and the orbit is slightly eccentric, then fragments of it may peel off at periastron passages.

\begin{deluxetable}{llcc}
\tablenum{8}
\tablecaption{Equilibrium Temperature on Orbit}
\label{tab:temps}
\tablewidth{0pt}
\tablehead{\multicolumn{2}{c}{Dependence$^a$} & \colhead{$T_{\rm eq}$(9.93 hr)} & \colhead{$T_{\rm eq}$(11.2 hr)}} 
\startdata
$T_{\rm BB}^b$   & $\simeq T_{\star} (R_{\star}/2d)^{1/2}$         & 385 K  & 370 K  \\
$T_{\rm hemi}^c$ & $\simeq T_{\star} (R_{\star}/\sqrt{2}d)^{1/2}$  & 458 K  & 440 K  \\
$T_{\rm ssp}^d$  & $\simeq T_{\star} (R_{\star}/d)^{1/2}$          & 545 K  & 525 K  \\
\multicolumn{2}{l}{$T_{\rm ssp}$ crater bottoms$^e$}               & 600 K  & 575 K  \\
$T_{\rm Rayl}^f$ & $\simeq T_{\star} (R_{\star}/2d)^{2/5}$         & 700 K  & 677 K  \\
\hline
$T_{\rm disk}^g$ & $\simeq \xi T_{\star} (R_{\star}/d)^{3/4} $     & 99 K   & 93 K \\
\enddata
\tablenotetext{a}{$R_{\star}$ and $T_{\star}$ are the white dwarf radius and temperature, and $d$ is the orbital distance of the heated body from the white dwarf.}
\tablenotetext{b}{$T_{\rm BB}$ is the blackbody temperature of a body that is small enough to be heated to a uniform temperature, but large compared to the wavelength of the incident radiation \citep[see][]{Xu_2018}.}
\tablenotetext{c}{Temperature of a body heated over a hemisphere to a uniform temperature over that hemisphere.}
\tablenotetext{d}{Temperature of a large body at its substellar point.}
\tablenotetext{e}{Crater bottoms near the substellar point may be heated to temperatures $\sim$10\% hotter than $T_{\rm ssp}$ (see \citealt{Hansen1977}; Keihm, S., personal communication).}
\tablenotetext{f}{Temperature of a particle much smaller than the wavelength of the incident radiation \citep[Rayleigh limit; see][]{Xu_2018}.}
\tablenotetext{g}{Temperature in the interior of a gas-dust disk \citep{Chiang1997} where the leading coefficient is $\xi =(2/3 \pi)^{1/4}$.  This entry below the horizontal line is the only one where shielding of the orbiting body by a disk is invoked.}

\end{deluxetable}

\begin{deluxetable}{lcc}
\tablenum{9}
\tablecaption{Asteroids vs.~Comets}
\label{tab:aster_comets}
\tablewidth{0pt}
\tablehead{\colhead{Parameter} & \colhead{Asteroid$^a$} & \colhead{Comet$^a$}} 
\startdata
Mass                      & up to $10^{24}$ g       &    up to $3 \times 10^{17}$ g   \\
Radius                    & up to 500 km            &    up to 30 km \\
Density                   & $1.0-3.5\,$g\,cm$^{-3}$ &  $0.3-0.6\,$g\,cm$^{-3}$ \\
Composition               & rocky                   & gas ices  \\
                          & minerals                & dust \\
Lifetime $@$ $10^9$ g/s   & $0-10$ Myr              & 10 days $-$ 10 yr    \\
\hline
Illustrative material     & Fe (1546 K)                  & H$_{2}$O (252 K)  \\
(Sublimation temp.)$^b$:  & SiO (1586 K)                 & CO$_{2}$ (194 K)  \\
                          & Fe$_{2}$SiO$_{4}$ (1585 K)   & CO (70 K) \\
                          & MgSiO$_{3}$ (1822 K)         & ... \\
                          & Mg$_2$SiO$_4$ (1904 K)       & ... \\
                          & SiO$_2$ (1977 K)             & ...  \\
                          & Al$_2$O$_3$ (1969 K)         & ... \\
                          & SiC (2098 K)                 & ... \\         
                          & C (2585 K)                   & ... \\
\enddata
\tablenotetext{a}{Generic properties of asteroids and comets taken from \citet{Carry12}.}
\tablenotetext{b}{Minerals and characteristic sublimation temperatures taken from \citet{vanLieshout2014}.}
\end{deluxetable}

\begin{figure}
	\epsscale{1.18}
	\plotone{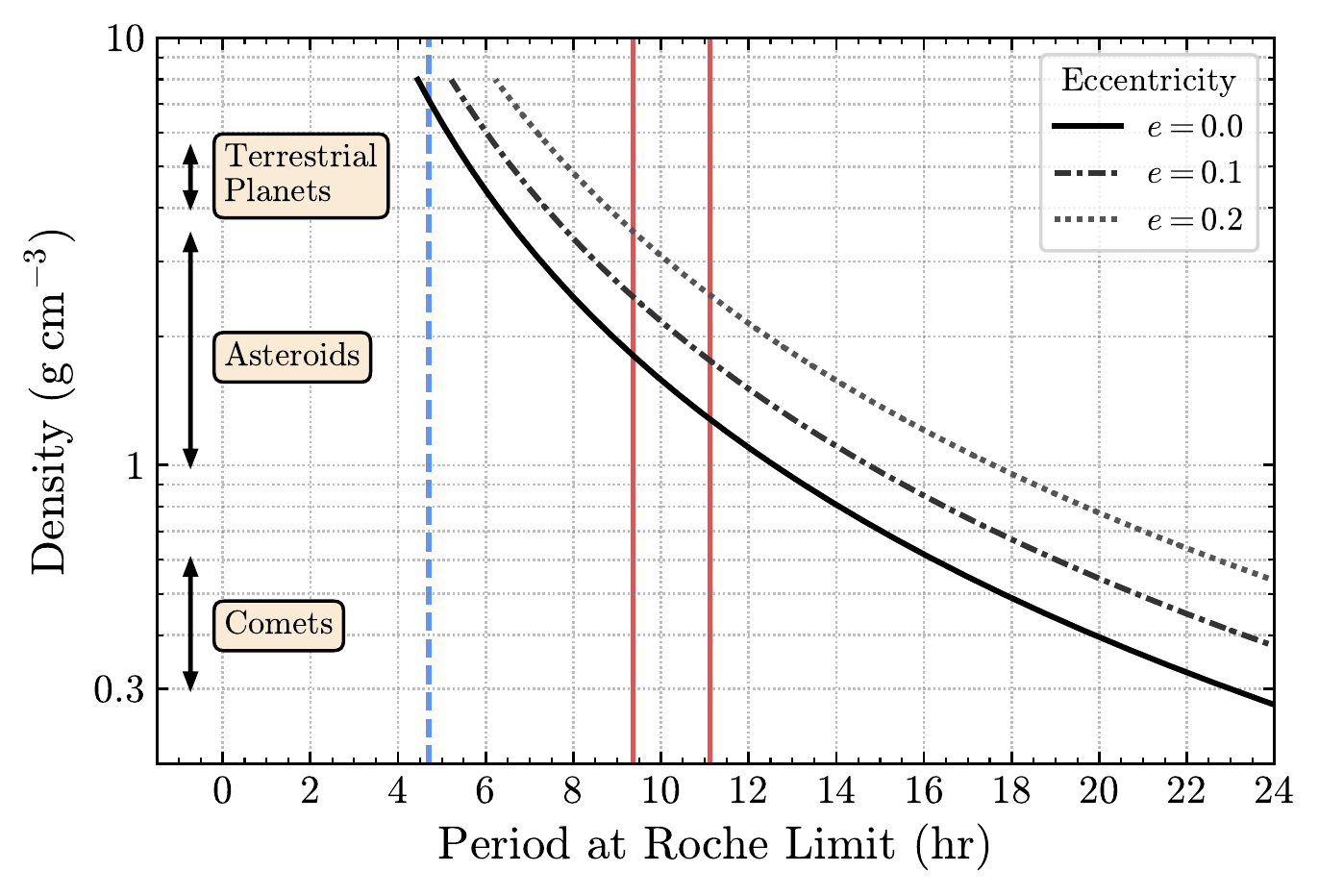}
    \caption{Minimum required mean density of the orbiting body as a function of its orbital period---described by Equation~(\ref{eqn:P_rho}). The typical density ranges for comets ($0.3{-}0.6\;\mathrm{g\,cm^{-3}}$), asteroids ($1.0{-}3.5\;\mathrm{g\,cm^{-3}}$), and terrestrial planets ($4.0{-}5.5\;\mathrm{g\,cm^{-3}}$) are indicated on the left-hand side. The dot-dashed and dotted curves are approximate modifications to Equation~(\ref{eqn:P_rho}) multiplied by a factor of $(1-e)^{-3/2}$ for eccentricities of 0.1 and 0.2, respectively, which illustrate how higher eccentricities at fixed periods push the required density higher. Vertical red lines mark the two periods identified in \obj while the vertical blue line is a characteristic period of WD\,1145+017. \label{fig:P_rho}}
\end{figure}

As it regards such close-in orbiting bodies, the original Roche limit \citep{Roche_1849} can be recast as an expression for the minimum allowed orbital period as a function of the mean density of the orbiting body
\begin{equation}
    P_{\rm min} \simeq \sqrt{\frac{3 \pi (2.44)^3}{G \bar{\rho}}} \simeq 12.6 \left(\frac{{\rm g~cm}^{-3}}{\bar{\rho}}\right)^{1/2}~{\rm hr}
    \label{eqn:P_rho}
\end{equation}
\citep[see e.g.][]{Rappaport2013,Rappaport2021}. A period of 9.93 hrs therefore requires a minimum mean density of only 1.6 $\mathrm{g \,cm}^{-3}$. By contrast, in the case of WD\,1145+017 with orbital periods near 4.5\,hr, the minimum mean density required is 7.8 $\mathrm{g \,cm}^{-3}$. This all assumes that the bodies are not held together by material forces. In the case of rubble piles (\citealt{Veras2014,Veras_2017_1}), the individual chunks of material are relatively free to peel off the host body when the size of the body exceeds its critical potential lobe. Equation~(\ref{eqn:P_rho}) is plotted in Figure~\ref{fig:P_rho} to help visualize where \obj and WD\,1145+017 fall with respect to the curve.

In terms of the material released by Roche breakup, we envision that smaller bodies might be produced in the following way. The orbiting bodies are likely naturally in slightly eccentric orbits. In that case the size of the critical potential lobe expands and contracts by a fractional amount $\pm
e$ around the orbit, where $e$ is its orbital eccentricity. In that case, the fractional volume of the rubble pile that is exposed to ejection of material at periastron passage is $\lesssim$ $3e$. Thus, for example, if $e = 10^{-4}$, then some $\lesssim 0.0003$ of the mass of the rubble pile is susceptible to breaking off each orbit. In that case, the rubble pile itself could have a lifetime of only a few thousand orbits, or only a few years.

Another promising mechanism for producing dust even in quite cool environments might be a set of enhanced cascading collisions. This scenario starts with a catastrophic collision between two substantial bodies which, after a number of subsequent orbits, produces an asymmetric debris disk with a region of high density that remains fixed near the original collision point \citep{Jackson2014}. Subsequent collisions within this high density region may be capable of producing fresh clouds of dust that last several orbits before spreading out in azimuth due to differential orbital velocities. While collisions provide a viable mechanism for producing dust, however, they have thus far only been examined theoretically in the contexts of debris disk morphology \citep{Jackson2014} and dust and gas disk production and accretion onto white dwarfs \citep{Kenyon_2017_1,Kenyon_2017_2}, and do not attempt to produce semi-persistent dust clouds that could yield the types of transits observed in WD\,1145+017 or \obj.

\subsubsection{Eccentric Debris Orbits}

Here we briefly consider some of the ramifications if the dust-emitting bodies are in substantially eccentric orbits. For a fixed observed orbital period, the semi-major axis of the orbit is fixed, regardless of the eccentricity.  The distance of closest approach, at which point the body is most vulnerable to tidal breakup, is $a (1-e)$, where $a$ is the semi-major axis of the orbit. 

One might guess that for eccentric orbits the equilibrium temperatures of a body would be higher than for a circular orbit of the same semi-major axis, due to the fact that the orbiting body spends time closer to the irradiating source. However, while the time-averaged stellar flux experienced by a body throughout an orbit increases with increasing eccentricity, the average distance also increases. The net result is an overall {\em decrease} in the time-averaged equilibrium temperature with increasing eccentricity \citep{Mendez2017}. As shown by \citet{Mendez2017}, for eccentricities of $e<0.5$, and assuming constant albedo throughout the orbit, the equilibrium temperature drops by only about 1\% or less compared to a circular orbit, and is lowered by no more than 10\% even out to eccentricities approaching unity. Thus, the equilibrium temperatures presented in Table~\ref{tab:temps} are upper limits in the case of circular orbits, but the effect of eccentric orbits on equilibrium temperatures is relatively small.

Another effect of eccentric orbits is that they change the $P_{\rm min}(\bar{\rho})$ relation give by Equation~(\ref{eqn:P_rho}). For an eccentric orbit we can roughly estimate\footnote{This estimate is based on the assumption that the linear size of the critical potential surface at periastron is proportional to the distance of closest approach.} that the minimum period in Equation~(\ref{eqn:P_rho}) is {\em increased} by a factor of $(1-e)^{-3/2}$; conversely, this increases the minimum mean density required to attain an orbit as short as is observed.  In particular, for $e=0.2$, the mean density for an object in a 9.93-hr orbit would become 3.1 g cm$^{-3}$ as opposed to 1.6 g cm$^{-3}$ for a circular orbit. The effect of eccentricity on the $P_{\rm min}(\bar{\rho})$ relation is illustrated graphically in Figure~\ref{fig:P_rho}.

A question may also arise about the stability of two nearby orbits where an eccentricity of one or both objects will lead to crossing orbits. To address this issue we have carried out a few numerical simulations with up to a half dozen low-mass bodies orbiting a white dwarf. Two of the orbits replicated those of the 9.93 and 11.2 hr orbits observed in \obj. We tested a set of orbits where all bodies had different semi-major axes, but one body had an eccentricity of 0.2, leading to crossing orbits. We also tested a set of orbits where two bodies have the {\em same} semi-major axis, but one is eccentric. For each simulation, all orbiting bodies were given the same mass, and we ran simulations with masses of $10^{20}$, $10^{23}$, $10^{26}$, and $10^{28}$\,g. The first two of these, where the orbiting bodies have masses comparable to Solar System asteroids, led to stable orbits for more than 3 years (i.e., longer than \obj has been observed), while the third of these, with masses comparable to the Moon, shows signs of perturbed orbits. For the last case, with masses roughly equal to Earth, the orbits were grossly unstable.

\subsection{Dust Required to Produce the Photometric Variations}

The minimum amount of dust that would be required to produce the photometric fluctuations that we see can be estimated as follows. The minimum total effective area that is blocked around the orbit by dust is given by
\begin{equation}
\mathcal{A} \gtrsim 4 \pi d R_{\star}f \tau
\label{eqn:area}
\end{equation}
where $d$ is the orbital radius of the dust emitting body, $R_{\star}$ is the white dwarf radius, $f$ is the fraction of the orbit where photometric modulation is observed, and $\tau$ is the optical depth of the dust in the visible band.  In the optically thin limit, the total cross section of all the dust particles, $\sigma_{\rm tot}$,  for a given total mass in dust, $M_{\rm dust}$ is
\begin{equation}
\sigma_{\rm tot} \simeq N_{\rm grain} \sigma_{\rm grain} \simeq \frac{3 M_{\rm dust}\sigma_{\rm grain}}{4 \pi \rho_d s^3}
\label{eqn:sigma}
\end{equation}
where $N_{\rm grain}$ is the total number of dust grains blocking light from the WD, $\sigma_{\rm grain}$ is the cross section of a single dust grain, $\rho_d$ is the mean bulk density of the dust particles, and $s$ is the mean effective size of a dust grain. We can write a general expression for the minimum required mass in dust by equating Equations~(\ref{eqn:area}) and (\ref{eqn:sigma}).  We then find:
\begin{equation}
M_{\rm dust} \gtrsim \frac{16 \pi d s R_{\star} f \tau \rho_d}{3 (\sigma_{\rm grain}/\sigma_{\rm geom})}
\label{eqn:mdust1}
\end{equation}
where the grain cross section is expressed in terms of its geometric cross section.  The most efficient grain scattering cross sections occur near $\sim$1 micron size particles, where $\sigma_{\rm grain}$ is still $\approx \sigma_{\rm geom}$.  Thus, we have 
\begin{equation}
M_{\rm dust} \gtrsim 16 d s R_{\star} \rho_d f \tau ~\simeq 10^{16}  \left(\frac{d}{2\,R_\odot}\right) \left(\frac{f\tau}{0.025}\right)~\mathrm{g}
\label{eqn:mdust2}
\end{equation}
and we have taken $s \simeq 1 \,\mu$m and $\rho_d \simeq 3$ g/cc.
Finally, if we somewhat arbitrarily assign a lifetime for the dust-grains of about 1 month (e.g., the timescale for the photometric modulations to undergo substantial change, then this leads to an inferred grain production rate of $\dot M_{\rm dust} \simeq 5 \times 10^9$ g s$^{-1}$. Similar approaches to estimating dust mass and loss rates have been discussed in, for example, \citet{Chen_Jura_2001}, \citet{Rappaport2012}, \citet{Vanderburg2015}, and numerous other papers.


We have limited information about the optical depths of the \obj dust clouds. A dip with depth of 10\% can be accounted for by either a cloud larger than the WD disk with an optical depth of 10\%, or an optically thick cloud that obstructs 10\% of the WD disk -- or some combination of these two extreme cases. The optically thick case will exhibit dip depths that are the same at all wavelengths. The optically thin case can exhibit anything between no depth dependence upon wavelength and a significant decrease of depth with increasing wavelength. Any observed wavelength dependence requires that the particle size distribution (PSD) include a significant component that is smaller than optical wavelengths (in addition to requiring optical depth $\lesssim 1$). For WD\,1145+017 it has been predicted \citep{Xu_2018} that the PSD should not include such small particles. Such particles are expected to become hotter than the volatile temperatures for typical materials because their emissivity is small at the particles' black body wavelength while they absorb starlight as well as large particles. 

The orbiting debris in \obj is far enough from its white dwarf that no such evaporation of small particles occurs, so we can expect to find that the \obj dust clouds consist of the full spectrum of the PSD. The most relevant observations we have for constraining wavelength dependence of the dip depth is the ZTF set of $g$ and $r$ band measurements (shown in Figure~\ref{fig:allphot}). Using 30-bin phased light curves in each filter, we find that the ZTF $g$ and $r$ band fluxes, when plotted against each other, have a slope of $(g-\langle g \rangle) \simeq  (0.38 \pm 0.10) (r-\langle r \rangle).$ This is contrary to what would be expected if the dust clouds included a significant component of small particles ($<$0.3 micron radius). We note, however, that the ZTF $g$ and $r$ band measurements are not made simultaneously and cover a broad time baseline, which may impact the observed amplitude correlation. Simultaneous multi-band observations would be ideal for addressing this important question.

\section{Summary \& Conclusions} \label{sec:conclusions}

In this work we have presented new and archival observations which confirm \obj as the third white dwarf to exhibit recurring transits caused by orbiting debris clouds. Photometry spanning more than 900 days was taken from \tess Sectors 4 and 31, as well as ground-based observations from ZTF, McDonald Observatory, SAAO, HAO, and JBO. The periods underlying the dusty transit behavior are 9.93\,hr (A-period) and 11.2\,hr (B-period), with the former periodicity being dominant. The peaks in a coherent Lomb-Scargle transform of the photometric data are about three times wider than would have been found for constant periods and stationary orbital modulation profiles. At this point we cannot determine whether pure amplitude and shape modulations are responsible for the broad peaks, or if slightly varying periods or the presence of multiple closely spaced periods for several transiting dust clouds are also responsible.

These periods are roughly twice as long as those observed in WD\,1145+017, the first of the white dwarfs found to be transited by multiple dusty bodies \citep{Vanderburg2015}, but still about two orders-of-magnitude shorter compared to the ${\approx}100$-day period observed in ZTF\,J0139+5245 \citep{Vanderbosch2020}. The nearly periodic dips in flux seen in WD\,1145+017 differ in two somewhat subtle ways from those observed thus far in \obj. First, the dips seen in WD\,1145+017 tend to be more localized with substantial portions of the orbital cycle where the flux level is constant between dips. In \obj there is variability throughout essentially the entire periodic cycle. Second, the modulation depths in WD\,1145+017 can reach 60\% of the flux. Thus far, the modulation amplitudes in \obj are about 10\% or less.

The existence of two periods argues strongly in favor of multiple dust-emitting orbiting bodies. In this regard, \obj is similar to WD\,1145+017, which also has multiple periodicities. The two periods in \obj, plus the changing modulation profiles exhibiting highly non-sinusoidal variations, argue against the periodicity being due to a white dwarf rotational modulation. If rotational modulation is involved, it would more likely be at the 11.2\,hr B-period since we have yet to identify any dip-like features with high-S/N follow-up observations. The B-period detections by \tess, however, suggest changes to both the period and modulation profile over time, characteristics which are inconsistent with rotational modulation.

We have used both an LDT spectrum and SED photometry to carefully characterize the white dwarf atmospheric parameters. We find $M_{\star} = 0.731 \pm 0.023\,M_{\odot}$, $T_{\mathrm{eff}} = 7630 \pm 140\,$K, \calcium$=-9.55\pm0.12$, and \hydro$<-3.5$. With new spectroscopic observations from MIKE and SOAR, we have also detected narrow Na~D doublet features with velocities 21.4$\pm$1.0\,km~s$^{-1}$ blue-shifted with respect to atmospheric features, which we determine to be of circumstellar origin. These lines are weak relative to the photospheric Ca~H and K lines, so we suspect the impact of circumstellar absorption on our calculated white dwarf atmospheric parameters to be insignificant. A weak H$\alpha$ line which we believe to be of atmospheric origin is also detected, though it is poorly fit by our DZ model.

Finally, we would like to advocate for further long-term monitoring of this source. Due to its brightness, \obj is accessible to good amateur photometry with 16-in telescopes, as demonstrated by our HAO and JBO photometry. This will help resolve the issue of whether the two periods we have found are unique, whether they are long-term stable, and whether the dips have many slightly different periods, as is the case for WD\,1145+017. Additionally, continued high-resolution spectroscopic monitoring will be useful for monitoring the velocities of the detected circumstellar features. Velocity variations over time may help constrain the eccentricity of the orbiting gaseous debris. Lastly, mid-infrared observations with {\em JWST} will be useful to better constrain the presence of a dusty debris disk in this system.


\acknowledgments

We thank the anonymous referee for a detailed review which led to significant improvements of this paper. We also thank Rob Robinson (Univ.~Texas, Austin) for helpful comments related to the properties of cataclysmic variables, and we thank Mathieu Choukroun (Caltech/JPL), Stephen J. Keihm (Caltech/JPL), Vishnu Reddy (Univ. Arizona) and Richard Binzel (MIT) for helpful discussions about solar system asteroids. We also thank Erik Dennihy (Gemini South) for an independent look at both the SED and SOAR spectra, and thank John Kuehne, Brian Roman, Coyne Gibson, Dave Doss, Anita Cochran, and all other McDonald observing support staff for making remote observing on the McDonald 2.1-m telescope possible during COVID-related travel restrictions.

Z.P.V. acknowledges support from the Wootton Center for Astrophysical Plasma Properties under U.S. Department of Energy cooperative agreement number DE-NA0003843. S.B. acknowledges support from the Laboratory Directed Research and Development program of Los Alamos National Laboratory under project number 20190624PRD2. C.M.\ and B.Z.\ acknowledge support from NSF grants SPG-1826583 and SPG-1826550. B.L.K. was supported by the M. Hildred Blewett Fellowship of the American Physical Society. J.J.H and T.M.H. acknowledge support from the National Science Foundation under Grant No. AST-1908119.

This paper employs data collected by the {\em Transiting Exoplanet Survey Satellite (\tess)} mission. Funding for the \tess mission is provided by the NASA's Science Mission Directorate.
This research was made possible through the use of the AAVSO Photometric All-Sky Survey (APASS), funded by the Robert Martin Ayers Sciences Fund and NSF AST-1412587.
This work has made use of data from the European Space Agency (ESA) mission
{\it Gaia} (\url{https://www.cosmos.esa.int/gaia}), processed by the {\it Gaia}
Data Processing and Analysis Consortium (DPAC,
\url{https://www.cosmos.esa.int/web/gaia/dpac/consortium}). Funding for the DPAC
has been provided by national institutions, in particular the institutions
participating in the {\it Gaia} Multilateral Agreement.
This work has also made use of observations obtained with the Samuel Oschin 48-inch Telescope at the Palomar Observatory as part of the Zwicky Transient Facility (ZTF) project. ZTF is supported by the NSF under Grant No. AST-1440341 and collaborating institutions.

Observations obtained using the 2.1\,m Otto Struve Telescope at the McDonald Observatory operated by The University of Texas at Austin are included in this work. 
Observations made using the Goodman Spectrograph mounted on the Southern Astrophysical Research (SOAR) telescope are also used in this work. 
We have also made use of observations obtained using the DeVeny Spectrograph mounted on the Lowell Discovery Telescope (LDT) at Lowell Observatory. 
This paper uses observations made at the South African Astronomical Observatory (SAAO).

\vspace{5mm}
\facilities{TESS, PO:1.2m (ZTF), Gaia, McD:Struve (ProEM), DCT (DeVeny), Magellan:Clay (MIKE), SOAR (Goodman), SAAO, ADS, CDS.
}

\software{Astropy \citep{Astropy_2013,Astropy_2018},
          {\sc iraf} (National Optical Astronomy Observatories),
          LMFIT \citep{LMFIT_2014},
          {\sc phot2lc} (Vanderbosch et al. in prep.),
          Photutils \citep{Bradley2020}, Starlink \citep{Starlink}, 
          CDS's (Strasbourg, France) SIMBAD and VizieR online pages and tables,
          and the NASA Astrophysics Data System (ADS) repositories.
          }
\\

\bibliography{ref}{}
\bibliographystyle{aasjournal}

\end{document}